\documentclass[floatfix,groupedaddress,twocolumn,showpacs,amsmath,pra,aps,amssymb,longbibliography,nofootinbib]{revtex4-2}

\usepackage{graphicx}
\usepackage{xcolor}
\usepackage{lipsum}
\usepackage{overpic}
\usepackage{hyperref}
\usepackage{txfonts}
\usepackage{amsmath,amsfonts,amssymb,verbatim,float,mathtools}
\usepackage{physics}
\usepackage[titletoc,title]{appendix} 
\usepackage{soul}
\usepackage{mathtools}
\hypersetup{
    unicode=true, 
    plainpages=false,
    colorlinks=true,
    linkcolor=blue,
    citecolor=blue,
    filecolor=blue,
    urlcolor=blue
}
\usepackage{bbold}

\newcommand{\cbE}{\boldsymbol{\mathbf{\cal E}}}
\newcommand{\unitvec}[1]{\hat{\mathbf{{#1}}}}
\renewcommand{\vec}[1]{\mathbf{#1}}

\newcommand\widthscale{1.0}

\begin{document}		
	\title{Spontaneous symmetry breaking in frustrated triangular atom arrays\\ due to cooperative
	light scattering}
	
	\author{C. D. Parmee}
	\affiliation{Department of Physics, Lancaster University, Lancaster, LA1 4YB, United Kingdom}
	\author{K. E. Ballantine}
	\affiliation{Department of Physics, Lancaster University, Lancaster, LA1 4YB, United Kingdom}
	\author{J. Ruostekoski}
	\affiliation{Department of Physics, Lancaster University, Lancaster, LA1 4YB, United Kingdom}
	
\begin{abstract}
We demonstrate the presence of an optical phase transition with frustration-induced spontaneous symmetry breaking in a triangular planar atomic array due to cooperative light-mediated interactions.
We show how the array geometry of triangle unit cells at low light intensities leads to degenerate collective radiative excitations forming nearly flat bands.
We drive degenerate pairs of collective excitations to be equally populated in both cases of the atomic polarization in the lattice plane and perpendicular to it. 
At higher intensities, above specific threshold values, this symmetry in the populations is spontaneously broken. 
We also develop an effective few-mode model that provides semianalytic descriptions of the symmetry-breaking threshold and infinite-lattice limit phase transition.
Surprisingly, we find how excitations due to dipolar interactions correspond to optical analogs of those found in frustrated magnets and superfluids, with closely related symmetry-breaking mechanisms despite the significant physical differences between these systems, opening potential for simulating even quantum magnetism. 
Transmitted light through the array conveys information about symmetry breaking in the hysteresis behavior of the spectrum.
Moreover, in a Mott-insulator state, the atomic positions are subject to zero-point quantum fluctuations. Interpreting each stochastic realization as a light-induced quantum measurement of the atomic position configuration, we find how strong nonlinearities and even weak position uncertainties lead to considerable measurement-induced symmetry breaking, while ensemble-averaging over many realizations restores the original symmetry and the unbroken state. Larger position uncertainty results in the formation of domains of different broken symmetries.
\end{abstract}

	\date{\today} 
	
	\maketitle

\section{Introduction}
		
Phase transitions are ubiquitous throughout physics and are signified by a sharp change in the behavior of a system with only a small modification in the system parameters.
Most phase transitions are described by the paradigm of spontaneous symmetry breaking (SSB), where the system configuration changes from one that respects the symmetry of the Hamiltonian to one that violates it.
There are many notable examples of SSB in a wide range of physical models. One of the best known is magnetic systems, where a random homogeneous configuration of magnetic moments spontaneously aligns below the Curie temperature. 
SSB is also found in magnetic systems with geometric frustration where, e.g., spins in a lattice experience competition between their interactions with one another and the lattice geometry~\cite{Moessner2006}.
In a triangular lattice with antiferromagnetic interactions, two spins in a triangle can anti-align, but the third cannot simultaneously anti-align with both. This results in degeneracy of the ground state, where the system spontaneously chooses between equally possible configurations, which leads to nontrivial magnetic behavior in closed systems~\cite{Wannier1950,Miyashita1984,Lee1984,Nishimori1988,Gekht1989,Chubukov1991}, or a multitude of steady-state phases in open systems~\cite{Li2021}.
Experimental implementations of closed classical frustrated spin systems on triangular lattices have been realized using even cold gases~\cite{Struck2019}, trapped ions~\cite{Kim2010,Britton2012} and coupled lasers~\cite{Nixon2013}.

Other notable examples of SSB can be found in optical systems of driven cold atom clouds that self-organize into one of many degenerate spatial structures due to phase transitions and SSB of translational symmetry. This phenomenon has been studied for atoms in single~\cite{Domokos2002,Black03,Asboth2005,Niedenzu2013,Lee14,Baumann2010,Caballero-Benitez2015} and multi-mode~\cite{Gopalakrishnan09,Vaidya18} cavities, where SSB is signaled by an increase in the intracavity photon number, and also in cold gas clouds with optomechanical and measurement back-action~\cite{Labeyrie2014,Robb15,Ivanov2020NAT,Baio21}.
More recently, optical phase transitions have been predicted for atoms trapped in an array without the presence of a cavity or back-action~\cite{Parmee2020}.

Atoms trapped in regular arrays are a particularly attractive physical system to study strong light-mediated cooperative interactions. There has been considerable recent theoretical interest, e.g., in the studies of optical responses in the limit of low light intensity (LLI)~\cite{Jenkins2012a,Bettles2016,Facchinetti16,Yoo2016,Jen17,Shahmoon,Sutherland1D,Perczel2017,Bettles2017,Facchinetti18,Asenjo_prx,Grankin18,Guimond2019,ballantine2020,Javanainen19,Needham19,Ballantine20Huygens,Alaee20,Yoo20,Shahmoon19,Ballantine21quantum,Rubies22,Ballantine21bilayer,Ballantine21PT}.
However, for stronger intensities of the incident field, nonlinear and quantum effects may become relevant~\cite{Kramer2016,Olmos16,Zhang2018,Henriet2018,Qu19,Ritsch_subr,williamson2020b,Cidrim20,Masson20,Parmee2020,Bettles2020,Orioli19,Williamson2020,Zhang20,Holzinger21,Zhang2022,Cosimo2021,Moreno2021,Panyella2022} and the collective effects are expected to result in novel light-induced phase transitions with clear observable signatures in the scattered light~\cite{Parmee2020}, and even the emergence of optical bistability~\cite{Parmee2021}. Near-resonance light scattering from cold atomic ensembles, including planar arrays, has also been experimentally studied at high atom densities where the collective effects are prominent~\cite{Rui2020, Balik2013,ChabeEtAlPRA2014,Jennewein_trans, Jenkins_thermshift,Dalibard_slab,Saint-Jalm2018,Machluf2018,Ferioli21}.

In this paper, we show the existence of an optical phase transition with frustration-induced SSB in an array of radiatively strongly coupled atoms. The atoms are trapped in a triangular array of triangle unit cells, and we find the lattice geometry results in remarkably flat bands and degenerate collective radiative excitations in the limit of LLI. However, for stronger driving, when the light intensity exceeds a threshold value, the symmetry is spontaneously broken, and one of the modes becomes dominant.
In addition to the numerical atom-by-atom simulations, we show this by developing a few-mode effective model that is valid in the large lattice limit where the number of relevant modes is reduced due to the phase-matching. 
The effective model provides qualitative understanding of the process of SSB, the threshold values for its emergence, and the infinite-lattice limit description of the phase transition.
Signatures of the SSB in the transmitted light from the ensemble are also identified, where we find sharp Fano-like resonances and hysteresis in the coherent transmission upon sweeping the laser frequency.
Excitation eigenmodes perpendicular to the array bear a notable resemblance to the spiral ground states in frustrated magnetic systems
with either positive or negative chirality and could even act as optical simulators for the physics of magnetic systems.
Surprisingly close analogies of phase transitions and SSB are found with other entirely different physical systems of triangular symmetries, such as Bose-condensed atoms with only contact interactions in a triangular optical lattice where the local phase of the superfluid acts as the spin excitation~\cite{Struck2019}. 

When the atoms in a Mott-insulator state are confined in the lowest vibrational levels of the lattice sites, as in recent light transmission experiments~\cite{Rui2020}, their positions are subject to zero-point quantum fluctuations. We solve the optical response by stochastic electrodynamics simulations, unraveling the atomic distribution into stochastic realizations of atomic positions~\cite{Lee16,Jenkins2012a}. Each stochastic realization leads to a characteristic response that is generated by the quantum fluctuations of the atomic positions. Individual stochastic configurations represent spontaneous symmetry breaking of atomic population dynamics and an optical phase transition.
In the presence of more significant fluctuations, the symmetry is broken differently across the lattice, resulting in the formation of domains. Ensemble-averaging over many stochastic realizations restores the original symmetry and the unbroken state.
Due to nonlinearities, even very weak position uncertainties lead to strongly enhanced changes in the atom response. We discuss the relationship between the stochastic electrodynamics and quantum trajectories and the interpretation of stochastic realizations as a light-induced quantum measurement of the atomic position configuration, resulting in measurement-induced symmetry breaking and phase transitions.

\section{Model}
\label{sec:setup}

\subsection{Atoms and light fields}
\begin{figure}
	\hspace*{0cm}
	\includegraphics[width=\columnwidth]{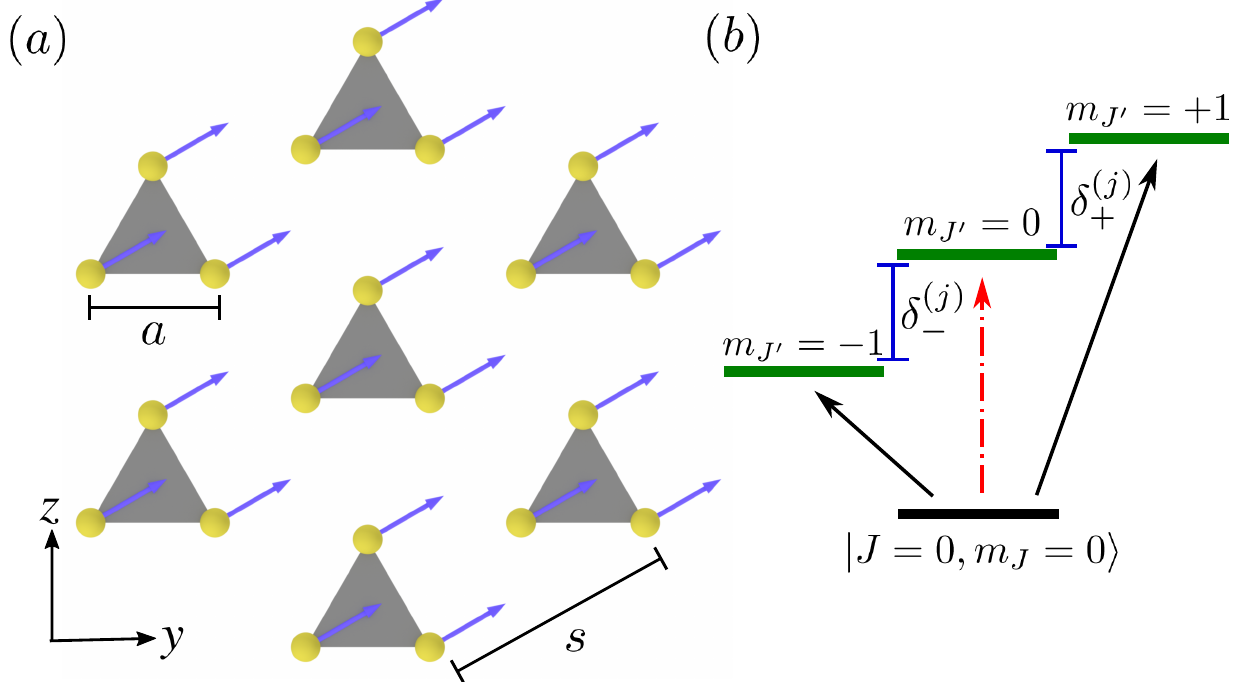}
	\vspace{-0.5cm}
	\caption{
		A planar array of atoms driven by incident light. (a) The atoms are trapped in a regular array of triangular unit cells with spacing $s$, where each unit cell is composed of three atoms with separation $a$. (b) The light is transmitted perpendicular to the array, driving a $\ket{J=0, m_J=0}\rightarrow\ket{J' = 1, m_{J'}=\mu}$ atomic transition with $\mu=\pm1$ for $y$-polarized light, or $\mu=0$ for $z$-polarized light. The $J' = 1$ manifold of atom $j$ has controllable level shifts $\delta_{\mu}^{(j)}$.
	}
	\label{Fig:Model}
\end{figure}

We consider a system of $N$ cold atoms trapped in a two-dimensional (2D) optical lattice, formed of a triangular array of unit cells in the $yz$ plane. Each unit cell is an equilateral triangle of three atoms with spacing $a$ [Fig.~\ref{Fig:Model}(a)], and a spacing $s$ between unit cells. 
The array is illuminated by an incident field which drives a $\ket{J=0, m_J=0}\rightarrow\ket{J' = 1, m_{J'}=\mu}$ atomic transition [Fig.~\ref{Fig:Model}(b)] and induces a dipole moment
$\textbf{d}_j = \mathcal{D}\sum_{\mu}\hat{\textbf{e}}_{\mu}\hat{\sigma}_{j\mu}^{-}$ on atom $j$, where $\mathcal{D}$ is the reduced dipole matrix element, $\ket{g}_j$ the ground state, $\ket{\mu}_j$ the excited states and $\hat{\sigma}_{j\mu}^{-}=|g\rangle_{j}\mbox{}_{j}\langle \mu|=(\hat{\sigma}_{j\mu}^+)^{\dagger}$.
For most of the paper, we consider the atoms fixed at the coordinates defined by the lattice. However, we also study position fluctuations in some cases by stochastically sampling the atomic positions $\{{\bf r}_1,\ldots,{\bf r}_N\}$ in the lattice sites~\cite{Jenkins2012a}.
By ignoring quantum fluctuations between different atoms~\cite{Lee16,Bettles2020}, the dynamical evolution of the system for any particular position configuration is determined by the following nonlinear equations
\begin{subequations}\label{Eq:SpinEquations}
	\begin{align}
	\dot{\rho}^{(j)}_{g \mu}=&[\text{i}\Delta^{(j)}-\text{i}\mu\delta_{\mu}^{(j)}-\gamma]\rho^{(j)}_{g \mu}+\text{i}\tilde{\mathcal{R}}_{\mu}^{(j)}\rho_{gg}^{(j)}-\text{i}\sum_{\alpha}\tilde{\mathcal{R}}_{\alpha}^{(j)}\rho_{\mu \alpha}^{(j)},\\
	\dot{\rho}^{(j)}_{\mu \nu}=&[\text{i}\mu\delta_{\mu}^{(j)}-\text{i}\nu\delta_{\nu}^{(j)}-2\gamma]{\rho}^{(j)}_{\mu \nu}+\text{i}\tilde{\mathcal{R}}_{\nu}^{(j)}\rho_{g\mu}^{(j)} -\text{i}[\tilde{\mathcal{R}}_{\mu}^{(j)}\rho_{g\nu}^{(j)}]^*,
	\end{align}
\end{subequations}
where $\rho_{gg}^{(j)}= 1- \sum_{\mu}\rho_{\mu \mu}^{(j)}$ and $\gamma = \mathcal{D}^2 k^3/(6\pi\epsilon_0\hbar)$ is the single atom linewidth.
The terms $\rho_{\mu \nu}^{(j)}=\langle\hat{\sigma}{}^{\mu\nu}_{j}\rangle$, where $\hat{\sigma}_{j}^{\mu\nu}=|\mu\rangle_{j}\mbox{}_{j}\langle \nu|$, represent excited state populations ($\mu=\nu$), and coherences between different excited states ($\mu\neq\nu$), while $\rho_{g \mu}^{(j)}=\langle\hat{\sigma}{}^{-}_{j\mu}\rangle$ are coherences between the ground and excited states.

We introduce in Eqs.~\eqref{Eq:SpinEquations} the effective Rabi frequencies~\cite{Parmee2020}
\begin{equation}\label{Eq:EffectiveRabi}
\tilde{\mathcal{R}}_{\mu}^{(j)} = \mathcal{R}_{g\mu}^{(j)} + \sum_{l \neq j}^{}\frac{6\pi\gamma}{k^3}\hat{\textbf{e}}_{\mu}^*\cdot \left[\mathsf{G}(\textbf{r}_j-\textbf{r}_l)\hat{\textbf{e}}_{ \nu}\right]\rho_{g\nu}^{(l)},
\end{equation}  
where ${\cal R}_{g\mu}^{(j)}=\mathcal{D}\hat{\textbf{e}}_{\mu}^*\cdot\boldsymbol{\mathcal{E}}{}^+(\textbf{r}_j)/\hbar$ are the usual Rabi frequencies of the incident field, and the corresponding intensities are $I^{(j)}/I_{\rm sat}=2\sum_{\mu}|{\cal R}_{g\mu}^{(j)}/\gamma|^2$, where $I_{\rm sat}= 4\pi^2\hbar c \gamma/3\lambda^3$ is the saturation intensity. The second term in Eq.~\eqref{Eq:EffectiveRabi} gives the scattered field from all other atoms at positions ${\bf r}_l$ in the ensemble which drives the atom at ${\bf r}_j$, and describes light-mediated dipole-dipole interactions via the dipole radiation kernel $\mathsf{G}$ ~\cite{Jackson}. 
The dipole-dipole interactions result in recurrent light scattering between the atoms, and lead to the formation of light-induced classical correlations when the solutions to Eqs.~\eqref{Eq:SpinEquations} are stochastically averaged over all realizations of fixed atomic positions~\cite{Javanainen1999a,Lee16,Bettles2020} (see Appendix A). Therefore, Eqs.~\eqref{Eq:SpinEquations} go beyond the standard mean-field equations as even though quantum correlations factorize, classical correlations do not.

The incident field illuminating the atoms is near monochromatic and propagating in the $x$ direction, with a positive frequency component $ \boldsymbol{\mathcal{E}}{}^+(\textbf{r})=\mathcal{E}_0(y,z)\hat{\textbf{e}}e^{\text{i}kx}$, where $k \hat{\textbf{e}}_x$ is the wavevector, $\omega = ck$ the frequency, and $\mathcal{E}_0(y,z)$ the amplitude, which is either constant or has a Gaussian profile. Observables are expressed in terms of slowly varying field amplitudes and atomic variables, $\boldsymbol{\mathcal{E}}{}^+e^{i\omega t} \rightarrow \boldsymbol{\mathcal{E}}{}^+$ and $\hat{\sigma}_{j\mu}^{-}e^{i\omega t} \rightarrow \hat{\sigma}_{j\mu}^{-}$.
The laser frequency detuning from the $m_{J'}=0$ level with resonance frequency $\omega_0$ in Eqs.~\eqref{Eq:SpinEquations} is given by $\Delta = \omega - \omega_0$, and $\delta_{\mu}^{(j)}$ represent the shifts of the $m_{J'}=\mu$ level on atom $j$, with $\delta_{0}^{(j)}=0$. Varying level shifts are used to break the isotropy of the $J=0\rightarrow J^\prime=1$ transition causing the dipoles to rotate in the $xy$ plane~\cite{Facchinetti16}. A controllable Zeeman-level splitting to the $J'=1$ manifold could be achieved either by using magnetic fields, or ac Stark shifts of lasers or microwaves~\cite{gerbier_pra_2006}.

\subsection{Low light intensity}
In the limit of LLI, the optical response of the system can be described in terms of collective radiative excitation eigenmodes~\cite{Rusek96,JenkinsLongPRB,Jenkins_long16}. We will show how the mode degeneracies lead to SSB at higher intensities by first analyzing the LLI limit, where atoms occupy the ground state, and changes to the coherences $\rho_{g\mu}^{(j)}$ are linearly proportional to the incident light field amplitude, $\boldsymbol{\mathcal{E}}^+$. The system then behaves as a linear set of $N$ coupled electric dipoles~\cite{Morice1995a,Ruostekoski1997a,Javanainen1999a,Sokolov2011,Jenkins2012a,Lee16}, with Eqs.~\eqref{Eq:SpinEquations} simplifying to
\begin{align}\label{LLI}
\dot{\textbf{b}}=\text{i}(\mathcal{H}+\delta\mathcal{H})\textbf{b}+\textbf{f},
\end{align}
where $\textbf{b}_{3j+\mu-1} = \rho_{g\mu}^{(j)}$ and $\textbf{f}_{3j+\mu-1} = \text{i}\mathcal{R}^{(j)}_{g\mu}$. The matrix $\mathcal{H}$ contains the light-induced dipole-dipole interactions between the atoms, with diagonal elements ${\rm i}\gamma$ and off-diagonal elements 
\begin{align}\label{DipoleMatrix}
\mathcal{H}_{3j+\mu-1,3l+\nu-1}= \frac{6\pi\gamma}{k^3}\hat{\textbf{e}}_{\mu}^*\cdot \left[\mathsf{G}(\textbf{r}_j-\textbf{r}_l)\hat{\textbf{e}}_{ \nu}\right],
\end{align} 
while the diagonal matrix $\delta\mathcal{H}$ contains the laser detuning and level shifts of the atoms, with elements $\Delta^{(j)}-\mu\delta_{\mu}^{(j)}$.
The $3N$ eigenmodes $\textbf{v}_{n}$ of $\mathcal{H}$, labeled by a single index $n$, describe the collective radiative excitations of the system, with complex eigenvalues $\lambda_{n} = \delta_{n}+\text{i}\upsilon_{n}$, where $\delta_{n}$ and $\upsilon_{n}$ are the collective line shift (from the resonance of the isolated atom) and linewidth, respectively~\cite{Rusek96,JenkinsLongPRB,Jenkins2012a,Lee16,Jenkins_long16}. 
In the presence of non-uniform level shifts $\Delta^{(j)}-\mu\delta_{\mu}^{(j)}$, $\delta\mathcal{H}$ generates a coupling between the different eigenmodes $\textbf{v}_{n}$ of $\mathcal{H}$.
The occupation of an arbitrary eigenmode $\textbf{v}_{n}$ for polarization amplitudes ${\bf b}$ can be determined by~\cite{Facchinetti16}
\begin{align}\label{Eq:ModePopulation}
		L_{n}=\frac{\left|\textbf{v}_{n}^{T}\textbf{b}\right|^2}{\sum_{n'}\left|\textbf{v}_{n'}^{T}\textbf{b}\right|^2}.
\end{align}
We also use Eq.~\eqref{Eq:ModePopulation} to determine the eigenmode occupation in the full nonlinear dynamics described by Eqs.~\eqref{Eq:SpinEquations}.

To approximate the behavior of the lattice for large $N$ without boundary effects, we also consider the modes of an infinite array of three-atom unit cells. In this limit, $\mathcal{H}$ reduces to a $9\times9$ matrix $\mathcal{H}_{\bf{q}}$, for each wavevector ${\bf q}$, containing the Fourier transform of the dipole kernel~\cite{Perczel2017} and with eigenvalues $\lambda_{{\bf q};m} = \delta_{{\bf q};m}+\text{i}\upsilon_{{\bf q};m}$.
The corresponding eigenmodes are Bloch waves,
\begin{subequations}\label{Eq:BlochWaves}
	\begin{align}
			&{\bf v}^{\rm{even}}_{{\bf q};m}({\bf R}_p) = {\bf u}_{{\bf q};m}\cos(\textbf{q}\cdot\textbf{R}_p),\\
			&{\bf v}^{\rm{odd}}_{{\bf q};m}({\bf R}_p)={\bf u}_{{\bf q};m}\sin(\textbf{q}\cdot\textbf{R}_p),
	\end{align}
\end{subequations}
where ${\bf R}_p$ gives the center of the $p^{\rm th}$ unit cell. The reciprocal vectors of ${\bf R}_p$ are given by $\textbf{g}=b_1\textbf{g}_1+b_2\textbf{g}_2$ for integers $b_{1}$, $b_{2}$, where $\textbf{g}_1=4\pi/(\sqrt{3}s)\unitvec{y}$,  $\textbf{g}_2=(2\pi/s)(\unitvec{y}/\sqrt{3}+\unitvec{z})$, and $\textbf{g}\cdot\textbf{R}_p=2\pi c$ for integer $c$, with modes invariant under translations $\textbf{q}\rightarrow\textbf{q}+\textbf{g}$.
For each $\textbf{q}$, the nine vectors ${\bf u}_{{\bf q};m}$, labeled by the subscript $m$, give the polarization amplitudes of one unit cell. Three (six) have a polarization lying out of (in) the atomic plane, and $|{\bf u}_{{\bf q};m}| = \sqrt{2/P}$ except for $|{\bf u}_{\Gamma,M;m}| = \sqrt{1/P}$, with $\Gamma={\bf 0}$, $M=\textbf{g}_1/2$ the high-symmetry points of the Brillouin zone, and $P$ the number of unit cells. The atomic transition resonance wavelength defines the light cone at $|\textbf{q}|=k$, where modes outside the light cone cannot couple to free-space modes while conserving momentum and energy as $|\textbf{q}|>\omega/c$. This results in eigenmodes that are completely dark with $\upsilon_{{\bf q};m}=0$, e.g., modes at the corner of the Brillouin zone, $K=(\textbf{g}_1+\textbf{g}_2)/3$, which lie outside the light cone for $s<2\lambda/\sqrt{3}$.

\section{Low light intensity modes and band structure}\label{KyleLLI}

We now describe the LLI eigenmodes of a single isolated triangle, $\bar{\textbf{v}}_{n}$, and the infinite lattice, $\textbf{v}_{\textbf{q};m}$.
We write the $9$-component single-triangle eigenmodes by $3$-component vectors $\bar{\textbf{v}}_n^{\mu}$ for each polarization $\mu$, with vector components labeled by the site index, i.e., $\bar{v}^{\mu,(j)}_n$. The $j=2$ component represents the top atom of the triangle.
When describing the polarization, we use radial, tangential and out-of-plane components $\bar{{\bf v}}_{n}^{r}$, $\bar{{\bf v}}_{n}^{t}$ and $\bar{{\bf v}}_{n}^{x}$, respectively (see Appendix B).

\begin{figure}
	\hspace*{-0cm}
	\includegraphics[width=\widthscale\columnwidth]{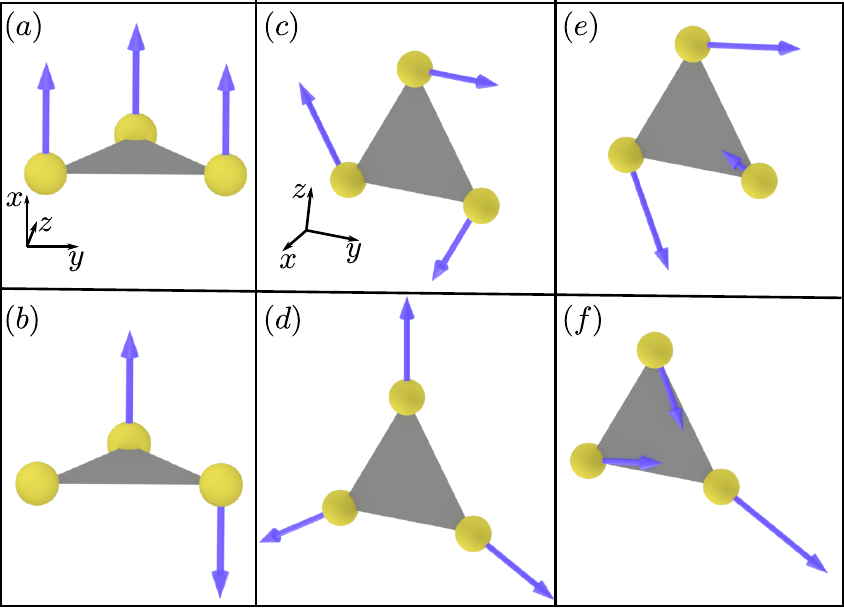}
	\vspace{-0.5cm}
	\caption{Radiative excitation eigenmodes of an isolated single triangle of atoms with spacing $a= 0.1\lambda$ in the limit of low light intensity. Real part of the polarization for the out-of-plane (a) uniform mode $\bar{\textbf{v}}_{\rm un}$ and (b) one of the two degenerate modes, $\bar{\textbf{v}}_1$; in-plane (c) tangential mode $\bar{\textbf{v}}_{\rm tan}$, (d) radial mode $\bar{\textbf{v}}_{\rm rad}$, and one of the two degenerate modes, (e) $\bar{\textbf{v}}_{a}$ and (f) $\bar{\textbf{v}}_{c}$. 
	Equivalent doubly-degenerate modes to (b,e,f) are formed by permuting the atomic polarization, with the third permutation a linear combination of the other two, and (e,f) have an unshown 16\% imaginary polarization component. 
	}
	\label{Fig:modes}
\end{figure}

\subsection{Single-triangle eigenmodes}

There are nine eigenmodes of an isolated single triangle [Fig.~\ref{Fig:modes}].  
Three modes have $\bar{\textbf{v}}_{n}^{r,t} = 0$, and a nonvanishing polarization perpendicular to the atomic plane, which we call the out-of-plane modes. One of these, $\bar{\textbf{v}}_{\rm un}$ [Fig.~\ref{Fig:modes}(a)], has a uniform phase profile and two of them, $\bar{\textbf{v}}_{1,2}$ [Fig.~\ref{Fig:modes}(b)], form a degenerate pair,
\begin{equation}\label{Eq:Eigenmodes}
\bar{\textbf{v}}_{\rm un}^x = \frac{1}{\sqrt{3}}\begin{pmatrix} 1 \\ 1 \\ 1 \end{pmatrix},\,
\bar{\textbf{v}}_{1}^x = \frac{1}{\sqrt{2}}\begin{pmatrix} 0 \\ -1 \\ +1 \end{pmatrix},
\,\bar{\textbf{v}}_{2}^x = \frac{1}{\sqrt{2}}\begin{pmatrix} -1 \\ +1 \\0 \end{pmatrix}.
\end{equation}
There are six eigenmodes with $\bar{\textbf{v}}_{n}^{x} = 0$ and a nonvanishing polarization in the atomic plane, which we call the in-plane modes. 
Two eigenmodes [Fig.~\ref{Fig:modes}(c,d)] have uniform radial (tangential) components, $\bar{\textbf{v}}_{\text{rad}}^{r} = (1,1,1)^{T}/\sqrt{3}$ [$\bar{\textbf{v}}_{\text{tan}}^{t} = (1,1,1)^{T}/\sqrt{3}$], and no tangential (radial) components, $\bar{\textbf{v}}_{\text{rad}}^{t} = 0$ [$\bar{\textbf{v}}_{\text{tan}}^{r} = 0$]. 
The remaining modes form two degenerate pairs, $\bar{\textbf{v}}_{a,b}$ and $\bar{\textbf{v}}_{c,d}$ [Fig.~\ref{Fig:modes}(e,f)], with radial and tangential components
\begin{subequations}\label{Eq:EigenmodesinPlaneAB}	
\begin{align}
&\bar{\textbf{v}}_{a/c}^{r} = \mathcal{N} \begin{pmatrix}\zeta_{\pm} \\ \zeta_{\pm}\\ -2\zeta_{\pm}\end{pmatrix},\
\bar{\textbf{v}}_{a/c}^{t} = \mathcal{N}\begin{pmatrix}\mp1\\ \pm1\\ 0\end{pmatrix}, \\
&\bar{\textbf{v}}_{b/d}^{r} = \mathcal{N}\begin{pmatrix}-2\zeta_{\pm}\\ \zeta_{\pm}\\ \zeta_{\pm}\end{pmatrix},\
\bar{\textbf{v}}_{b/d}^{t} = \mathcal{N}\begin{pmatrix}0\\ \mp1\\ \pm1\\ \end{pmatrix},
\end{align}
\end{subequations}
where $\mathcal{N} =(2+6\zeta_{\pm}^2)^{-1/2}$ and the parameter $\zeta_{\pm}$ is a function of the spacing $a$ (see Appendix B). 

The eigenmode degeneracies are caused by the geometry, which leads to a frustration-like effect. This is most easily explained by considering the out-of-plane modes, Eqs.~\eqref{Eq:Eigenmodes}, where three distinct polarization configurations can form on the triangle. The first is the fully symmetric mode with all the dipoles aligned, $\bar{\textbf{v}}_{\text{un}}$. The other two configurations have to be an anti-symmetric combination of the dipole moments, which can easily be formed for two dipoles by anti-aligning the moments. However, a third dipole cannot anti-align with the first two because of the geometry, and therefore must instead vanish. Because it can vanish from any of the three sites in the triangle, this results in a degeneracy as each configuration is equivalent with the same line shift and linewidth, e.g., the $\bar{\textbf{v}}_{1}$ and $\bar{\textbf{v}}_{2}$ modes. The degeneracy is only two-fold rather than three-fold as any one configuration can be formed by a linear combination of the other two.

\begin{figure}
	\hspace*{-0cm}
	\includegraphics[width=\widthscale\columnwidth]{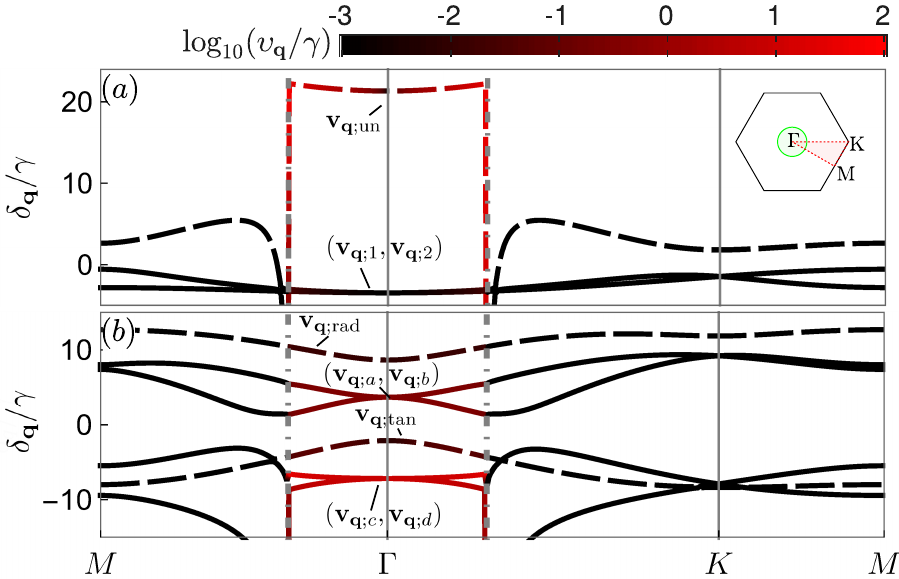}
	\vspace{-0.5cm}
	\caption{
	Collective line shifts $\delta_{\bf q}$ and linewidths $\upsilon_{\bf q}$ of low light intensity eigenmodes in an infinite triangular array of three-atom unit cells with the intra- and inter-unit-cell spacings $a=0.1\lambda$ and $s=2a$, respectively. The band structure is shown for the (a) three out-of-plane and (b) six in-plane modes at wavevectors ${\bf q}$ along the high-symmetry directions (inset). All the modes have a vanishing linewidth outside the light cone (gray dot-dashed line), and also inside the light cone at ${\bf q}=\Gamma$ for the out-of-plane, tangential and radial modes. Both the out-of-plane modes, ${\bf v}_{{\bf q};1}$ and ${\bf v}_{{\bf q};2}$, and in-plane modes, ${\bf v}_{{\bf q};a}$, ${\bf v}_{{\bf q};b}$, ${\bf v}_{{\bf q};c}$ and ${\bf v}_{{\bf q};d}$, become degenerate with their respective partner at $\Gamma$ and $K$. 
	}
	\label{Fig:Bandsxyz}
\end{figure}
\begin{figure}
	\hspace*{-0cm}
	\includegraphics[width=\widthscale\columnwidth]{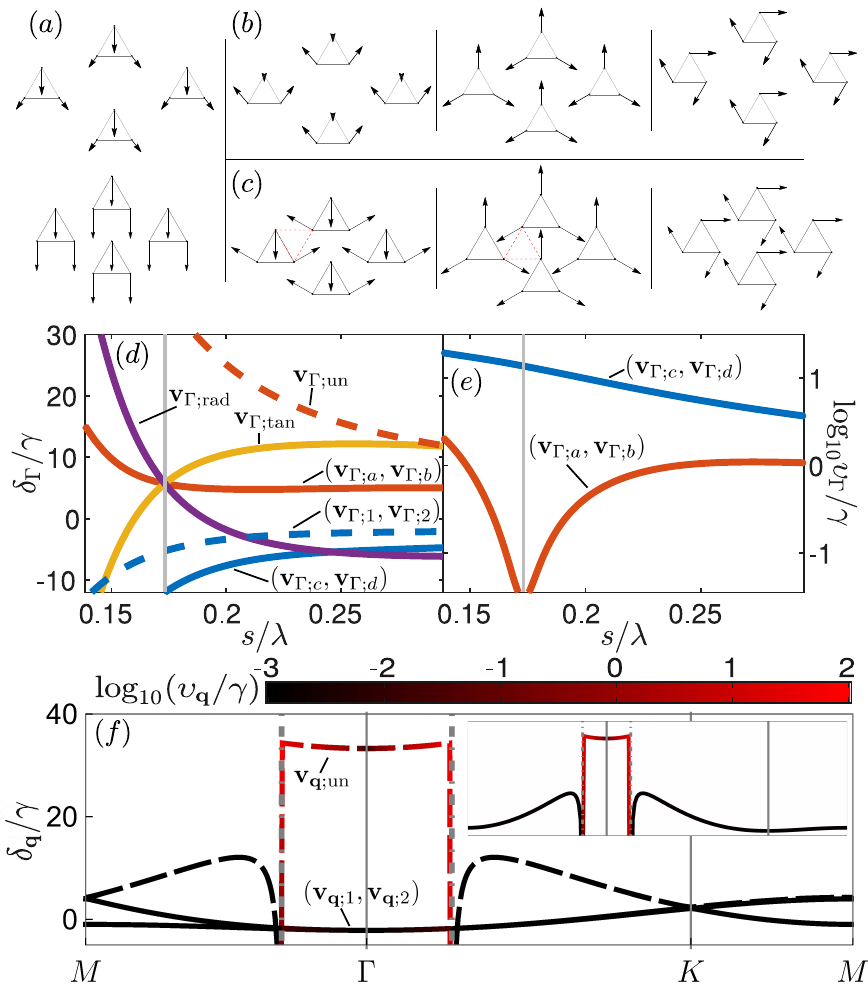}
	\vspace{-0.5cm}
    	\caption{Collective eigenmodes and band structures of a infinite triangular array of three-atom unit cells with the intra-unit-cell spacing $a=0.1\lambda$ and varying the inter-unit-cell spacing $s$. Real part of the atomic polarization for (a) $\vec{v}_{\Gamma;c}$ and $\vec{v}_{\Gamma;d}$, with $s=2\sqrt{3}a$ (top, lattice compressed for space) and $\sqrt{3}a$ (bottom), where the dipoles align to form a uniform mode, and (b) $\vec{v}_{\Gamma;a}$, $\vec{v}_{\Gamma;b}$, $\vec{v}_{\Gamma;{\rm rad}}$, $\vec{v}_{\Gamma;{\rm tan}}$ (left to right) for $s=2\sqrt{3}a$ and (c) $\sqrt{3}a$, where the modes become degenerate. Red dashed triangles illustrate equivalent choices of unit cells. For the degenerate pairs, one linear combination is shown, with unshown imaginary components comparable or smaller than in Fig.~\ref{Fig:modes}. 
    	Corresponding collective (d) line shifts and (e) linewidths of the modes in (a-c) as a function of $s$, showing the degeneracy at $s=\sqrt{3}a$ (gray vertical line). Note that $\upsilon_{\Gamma;1}$, $\upsilon_{\Gamma;2}$, $\upsilon_{\Gamma;{\rm un}}$, $\upsilon_{\Gamma;{\rm tan}}$, $\upsilon_{\Gamma;{\rm rad}}$ vanish for $s<\lambda$. 
        (f) Out-of-plane mode band structure for the standard triangular lattice, $s=\sqrt{3}a$. The $\Gamma$ point of the ${\bf v}_{{\bf q};1}$ and ${\bf v}_{{\bf q};2}$ bands corresponds to the $K$ point of the larger Bravais lattice Brillouin zone with spacing $a$ (inset). 
	}
	\label{Fig:VaryingS}
\end{figure}

\subsection{Infinite lattice eigenmodes}

We now analyze the eigenmodes of the infinite lattice, given by Bloch waves, Eq.~\eqref{Eq:BlochWaves}.
We first study lattices with $s$ notably larger than $\sqrt{3}a$, where interactions within a unit cell are stronger than those between neighboring unit cells. In this limit, the infinite lattice modes can be constructed from the single-triangle eigenmodes with $\textbf{u}_{\textbf{q};m} \approx \bar{\textbf{v}}_{m}$. 
Note that for the degenerate
modes, $\bar{\textbf{v}}_{m}$ may instead be a (${\bf q}$-dependent) linear combination of the modes defined in Eqs.~\eqref{Eq:Eigenmodes} and Eqs.~\eqref{Eq:EigenmodesinPlaneAB}.
The corresponding eigenmode band structure $\lambda_{\textbf{q}}$ is shown in Fig.~\ref{Fig:Bandsxyz}.
The bands formed from the degenerate single-triangle modes, $\bar{\textbf{v}}_{1,2}$, $\bar{\textbf{v}}_{a,b}$, and $\bar{\textbf{v}}_{c,d}$ remain distinct from their respective pair, except at the $\Gamma$ and $K$ points, where they become degenerate.
The degeneracy at the $\Gamma$ point is expected as the infinite lattice eigenmode is simply a uniform repetition of the single-triangle modes.
However, at the $K$ point, the reasoning is more subtle, with the degeneracy arising because swapping the eigenmode with its pair, e.g., swapping ${\bf v}_{K;1}$ and ${\bf v}_{K;2}$, is equivalent to a reflection $y\rightarrow -y$ under which the lattice, and hence also the eigenmodes, are invariant. 
Notably, several of the bands in Figs.~\ref{Fig:Bandsxyz} are relatively flat, especially close to the $\Gamma$ point, leading to a high density of modes with similar resonance frequencies. Such flat bands could have interesting optical applications and they are also specific for  the triangular lattice geometry; e.g., for a square unit cells we do not find similar flat bands.

Modes outside the light cone are completely dark, but the out-of-plane modes, and the radial and tangential in-plane modes, also have vanishing linewidth inside the light cone at the $\Gamma$ point.
This is because, for $s<\lambda$, light can only be emitted into the zeroth-order Bragg peak perpendicular to the array~\cite{CAIT,Facchinetti16,Javanainen19}, but for ${\bf v}_{\Gamma;1}$, ${\bf v}_{\Gamma;2}$ and ${\bf v}_{\Gamma;{\rm un}}$ the out-of-plane electric-dipole moments do not radiate in this direction.
While the radial and tangential modes feature in-plane electric-dipole moments, the polarization averages to zero on each unit cell, and they form effective out-of-plane electric quadrupoles and magnetic dipoles, respectively (similarly found in a regular array of square unit cells~\cite{Ballantine20Huygens}), which, again do not radiate along their axis.

We now study the eigenmodes in the limit $s\sim \sqrt{3}a$, when the spacing between nearest-neighbor atoms on different unit cells becomes equal to that within a unit cell, and the array is equivalent to a standard Bravais triangular lattice with spacing $a$. The approximation $\textbf{u}_{\textbf{q};m} \approx \bar{\textbf{v}}_{m}$ breaks down for some of the modes as interactions between atoms in different unit cells become important.
This is demonstrated in Fig.~\ref{Fig:VaryingS}(a) for the degenerate mode pair ${\bf v}_{\Gamma;c}$ and ${\bf v}_{\Gamma;d}$, where upon reducing $s$, the dipole moments in each unit cell start to align with one another, with the two smaller moments increasing in magnitude. 
At $s=\sqrt{3}a$, each mode ends up with a spatially uniform polarization, where the dipoles all have the same magnitude and direction in a given mode. 
The modes remain degenerate, and therefore we can choose a linear combination where the dipoles point along the $y$ or $z$ direction of the lattice, ${\bf v}_{\Gamma;y}$ and ${\bf v}_{\Gamma;z}$, respectively.

The polarization of the ${\bf v}_{\Gamma;a}$ and ${\bf v}_{\Gamma;b}$ pair also changes with $s$. We find that the $\textbf{v}_{\Gamma;a}$, $\textbf{v}_{\Gamma;b}$, $\textbf{v}_{\Gamma;{\rm rad}}$ and $\textbf{v}_{\Gamma;{\rm tan}}$ modes, which are distinct for $s>\sqrt{3}a$ [Fig.~\ref{Fig:VaryingS}(b)], become degenerate when $s=\sqrt{3}a$. This occurs because the decision to associate an atom with any particular two of its nearest neighbors becomes arbitrary. Then, for an alternative choice of unit cell, one linear combination of $\textbf{v}_{\Gamma;a}$ and $\textbf{v}_{\Gamma;b}$ has an equivalent polarization to $\textbf{v}_{\Gamma;{\rm rad}}$  [as shown in Fig.~\ref{Fig:VaryingS}(c)], and the other to $\textbf{v}_{\Gamma;{\rm tan}}$.
The degeneracy is also shown in the line shifts, Fig.~\ref{Fig:VaryingS}(d), and linewidths Fig.~\ref{Fig:VaryingS}(e). 

The polarization of the out-of-plane modes $\textbf{v}_{\Gamma;1}$, $\textbf{v}_{\Gamma;2}$ and $\textbf{v}_{\Gamma;{\rm un}}$ is unaffected by changes in $s$. 
For $s=\sqrt{3}a$, the band structure of the triangular array of unit cells can be mapped to the band structure of a standard triangular (Bravais) lattice with spacing $a$. 
In Fig.~\ref{Fig:VaryingS}(f), we show the unfolding of the band structure for the out-of-plane modes, where the ${\bf v}_{{\Gamma};1}$ and ${\bf v}_{\Gamma;2}$ modes map to the $K$ point in the larger Brillouin zone of the standard lattice, which is outside the light cone, and so these modes are completely dark.
The standard triangular atomic lattice has previously been studied for transmission~\cite{Bettles2016} and topological edge modes under an applied magnetic field~\cite{Perczel2017}.
As we will see in Sec.~\ref{sec:oop_symmetry}, the out-of-plane modes of the standard triangle lattice can also be identified as optical analogs for spin configurations found in frustrated magnetic systems.

\section{Spontaneous symmetry breaking}

In the previous section, we saw how the geometry of the lattice resulted in degeneracies of the single-triangle and lattice eigenmodes. We now show that increasing the incident light intensity results in a phase transition where SSB is caused by frustration and the population of one of the modes that are degenerate in the LLI limit becomes dominant.

\subsection{Out-of-plane spontaneous symmetry breaking}
\label{sec:oop_symmetry}

We demonstrate SSB in the mode occupation by exciting the out-of-plane degenerate eigenmode pair $\textbf{v}_{1}$ and $\textbf{v}_{2}$. These modes are the corresponding finite-lattice versions of $\textbf{v}_{\Gamma;1}$ and $\textbf{v}_{\Gamma;2}$, and have a uniform repetition of the single-triangle modes [Eqs.~\eqref{Eq:Eigenmodes}] with the same phase profile, but with a decreasing amplitude close to the lattice edge.
We consider a triangular array ($s=\sqrt{3}a$) with spacing $a = 0.1\lambda$. Small spacings are achievable using, e.g., the $J=0\rightarrow J^\prime=1$ $\lambda\approx 2.6\mu{\rm m}$ transition in Sr, with $a \simeq 0.08\lambda$~\cite{Olmos13,Ballantine22}.
The array hosts an in-plane LLI excitation eigenmode $\textbf{v}_{y}$, with all the atomic dipoles in phase pointing in the $y$-direction (Sec.~\ref{KyleLLI}). The mode is driven with a $y$-polarized Gaussian beam, achieving a closer overlap in a finite-sized array than a plane wave~\cite{Facchinetti16}.

\begin{figure}
	\hspace*{-0cm}
	\includegraphics[width=\columnwidth]{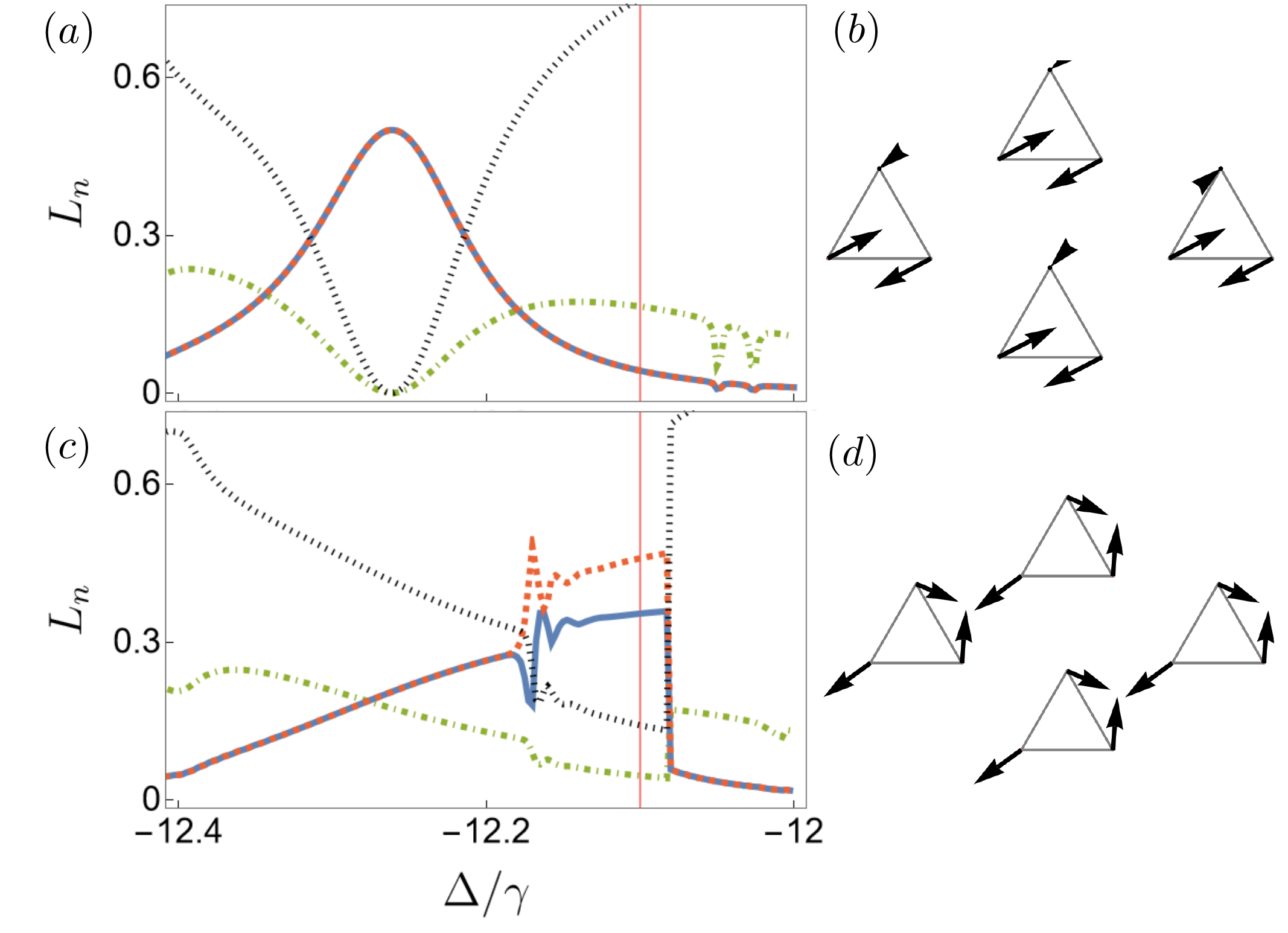}
	\vspace{-0.7cm}
	\caption{Spontaneous symmetry breaking shown by the atomic out-of-plane polarization density and mode occupation spectrum in a standard triangular array of $N=273$ atoms with spacing $a=0.1\lambda$ driven by a $y$-polarized Gaussian beam with width $w_0=0.8\lambda$. Level shifts [Eq.~\eqref{Eq:ZeemanShift} with $\alpha/\gamma= -0.17$] couple the light to out-of-plane excitations.
	(a,c) Occupations $L_n$ [Eq.~\eqref{Eq:ModePopulation}] of ${\bf v}_1$ (blue line), ${\bf v}_2$ (red dashed line) and ${\bf v}_y$ (green dot-dashed line), and sum of other eigenmode occupations (black dotted line) as a function of the laser frequency detuning from the atomic resonance $\Delta$. 
	(a) Unbroken symmetry with equal occupations in the LLI limit, while at the intensity (c) $\sum_j I^{(j)}/NI_{\rm{sat}}=0.37$, the mode symmetry is broken. (b,d) Corresponding change in the polarization $x$-component [length (angle) corresponds to dipole magnitude (phase)] shown for $\Delta/\gamma = -12.1$ (red line) which (b) vanishes on the top atom of each unit cell for unbroken symmetry, but is (d) nonvanishing for the broken symmetry.
	}
	\label{Fig:SymmetryBreakingOP}
\end{figure}

The out-of-plane modes $\textbf{v}_{1}$, $\textbf{v}_{2}$ are very subradiant and do not directly couple to the incident field. 
They can instead be populated by applying appropriate level shifts $\delta_{\pm}^{(j)}$ (see Fig.~\ref{Fig:Model} and Appendix C) with strength $\alpha$
that break the symmetry such that $\textbf{v}_{y}$, $\textbf{v}_{1}$, and $\textbf{v}_{2}$ are no longer eigenmodes of the collective light-matter system.
The level shifts couple $\textbf{v}_{y}$, driven by $y$-polarized incident field, to the out-of-plane modes by transferring the population to $\textbf{v}_{1}$, $\textbf{v}_{2}$.
The result is an equal occupation $L_1=L_2$ [Eq.~\eqref{Eq:ModePopulation}] of $\textbf{v}_{1}$ and $\textbf{v}_{2}$, as demonstrated in Fig.~\ref{Fig:SymmetryBreakingOP}(a) and maximal values $L_{1,2} \simeq 0.5$ at the eigenmode resonance, $\Delta/\gamma \simeq -12.3$. 
The corresponding final steady-state collective excitation [Fig.~\ref{Fig:SymmetryBreakingOP}(b)] has a vanishing $x$-polarization component on the top atom in each unit cell due to the equal superposition of $\textbf{v}_{1}$ and $\textbf{v}_{2}$.

At increasing incident light intensities, we find that the symmetry between mode populations spontaneously breaks. This is illustrated in Fig.~\ref{Fig:SymmetryBreakingOP}(c) for $L_2>L_1$ where a nonzero $x$-component of atomic polarization emerges on the top atom of each unit cell [Fig.~\ref{Fig:SymmetryBreakingOP}(d)].
The presence of this symmetry breaking is surprising because there are no terms in the dynamics [Eqs.~\eqref{Eq:SpinEquations}] that favor one mode occupation over the other. 
Instead, a nonzero population of the $J=1$ excited state angular momentum $x$-component (out-of-plane), 
\begin{align}\label{Eq:Cartesian}
	&\rho_{xx}^{(j)} = \frac{1}{2}[\rho_{++}^{(j)}+\rho_{--}^{(j)}-\rho_{+-}^{(j)}-\rho_{-+}^{(j)}],
\end{align}
emerges beyond the LLI limit [with $\rho_{++}^{(j)}$, etc., defined below Eqs.~\eqref{Eq:SpinEquations}], resulting in nonlinear mode interactions and frustration which breaks the symmetry.

In addition to the numerical simulations of the cooperative response of all the atoms, we analyze the system by developing a few-mode effective model that is valid in the large lattice limit when only a selective number of modes play a relevant role due to the phase-matching (Appendix D). The effective model illustrates the emergence of SSB, provides the description of the $N\rightarrow\infty$ phase transition, the SSB threshold behavior, and light transmission signal through the array.
The array polarization in the large lattice limit is
${\bf b} = \sqrt{N}[c_1{\bf v}_{\Gamma;1}+c_2{\bf v}_{\Gamma;2}+c_{y}{\bf v}_{\Gamma;y}]$ and
the steady-state mode amplitudes are given by (Appendix D)
\begin{subequations}\label{Eq:CSolnsMain}
	\begin{align}
	&c_{1,2} = \text{i}\frac{\alpha(P_t-1)[Z_- \pm 2\sqrt{3} P_m(\lambda_{d}-\text{i}\gamma)]}{Z_-[\alpha^2-Z_+Z_y]+4P_m^2(\lambda_d-\text{i}\gamma)Z_y}\mathcal{R}_{gy},\label{Eq:c12}\\
	&c_{y} = \frac{(P_t-1)[4P_m^2(\lambda_{d}-\text{i}\gamma)^2-Z_{+}Z_{-}]}{Z_-[\alpha^2-Z_+Z_y]+4P_m^2(\lambda_d-\text{i}\gamma)Z_y}\mathcal{R}_{gy},\label{Eq:cip}
	\end{align}
\end{subequations}
where
\begin{subequations}\label{Eq:Zdefn}
\begin{align}
&Z_{\pm}(\Delta)=\Delta +\delta_d(1-2P_{\pm})+\text{i}[\upsilon_d(1-2P_{\pm})+2\gamma P_{\pm}]\label{Eq:Zdefnd}, \\
&Z_{y}(\Delta)=\Delta +\delta_{y}(1-P_{t})+\text{i}[\upsilon_{y}(1-P_{t})+\gamma P_{t}]\label{Eq:Zdefny}.
\end{align}
\end{subequations}
The eigenstate projections $P_t = (P_{+}+P_{-})/2$,
\begin{align}\label{Eq:PDefinitions}
    &P_m = \sum_{j}v^{x,(j)}_{+}v^{x,(j)}_{-}\rho_{xx}^{(j)},\, P_{\pm}=\sum_{j}[v^{x,(j)}_{\pm}]^2\rho_{xx}^{(j)},
\end{align}
with $\textbf{v}_{\pm} = \mathcal{N}_{\pm}[\textbf{v}_{\Gamma;1}\pm \textbf{v}_{\Gamma;2}]$, expressed in Eq.~\eqref{Eq:PDefinitions} using the notation from the start of Sec.~\ref{KyleLLI}, and $\mathcal{N}_{+}=1$, $\mathcal{N}_{-}=1/\sqrt{3}$.
The $\textbf{v}_{\Gamma;1}$ and $\textbf{v}_{\Gamma;2}$ mode eigenvalue $\lambda_d = \delta_{d}+\text{i}\upsilon_d$, with $\delta_d = \delta_{1}= \delta_{2}$ and $\upsilon_d = \upsilon_{1}= \upsilon_{2}$, is very subradiant in our system (e.g., $\upsilon_{d}/\gamma\simeq 3\times 10^{-4}$, for $N=273$) with
$\lim_{N\rightarrow\infty}\upsilon_{d}=0$, as discussed in Sec.~\ref{KyleLLI}.
The main focus of Eqs.~\eqref{Eq:CSolnsMain} is $P_m$. 
$P_{\pm}$ shift the eigenmode resonances and broaden the linewidths [see Eqs.~\eqref{Eq:Zdefn}], but it is when $P_m\neq 0$ that $c_{1}\neq c_{2}$, therefore demonstrating that nonlinearity is essential for the SSB.
When the symmetry is unbroken ($P_m=0$), Eqs.~\eqref{Eq:CSolnsMain} are reminiscent of a three-level atom, with $\textbf{v}_{\Gamma;y}$ and the equal superposition of $\textbf{v}_{\Gamma;1}$ and $\textbf{v}_{\Gamma;2}$ analogs of the bright and dark states in electromagnetically induced transparency (EIT)~\cite{FleischhauerEtAlRMP2005}, respectively (see Appendices C and D). 
The level shifts couple these two states together, and a large transfer of the in-plane to out-of-plane mode population is achieved when $\alpha \gg \upsilon_{d}$ for both the symmetry-broken and unbroken case.

The dynamics in the large lattice limit confirm that the frustration breaks the symmetry spontaneously. $P_m$ obeys
\begin{equation}\label{Eq:PQuadratic}
	P_m(A_3+ A_2P_m^2 + A_1P_m^4) = 0,
\end{equation}
(see Appendix D for coefficients) which always has a solution preserving the symmetry, $P_m=0$, while nonzero values of $P_m$ are roots of the quartic in brackets. The signs of the nonzero solutions to Eq.~\eqref{Eq:PQuadratic} can be arbitrary, reflecting how either mode can be spontaneously favored when the symmetry breaks. 
This can be seen from Eq.~\eqref{Eq:c12}, where $P_m$ introduces opposite shifts in $c_{1}$ and $c_{2}$, and therefore the sign of $P_m$ results in either $|c_1|>|c_2|$ or $|c_1|<|c_2|$.
It is also observable in the finite-size dynamics, where altering the initial conditions from ${\bf b}=0$ to ${\bf b}=0.01{\bf v}_{1}$ (${\bf b}=0.01{\bf v}_{2}$) results in the dominantly occupied mode completely switching with $L_{1} > L_{2}$ ($L_{2} > L_{1}$) in the steady state.

This seeding of the final steady state also occurs due to fluctuations in the atom positions, which can be incorporated by accounting for the root-mean-square width $\sigma$ of the lattice site wave function (see Appendix A). In a Mott-insulator state, each lattice site can contain a single atom that is confined in a vibrational ground state, with the position determined by zero-point quantum fluctuations. Light transmission experiments in such a system demonstrated considerable resonance narrowing~\cite{Rui2020} due to cooperativity. 

\begin{figure}
	\hspace*{-0cm}
	\includegraphics[width=\widthscale\columnwidth]{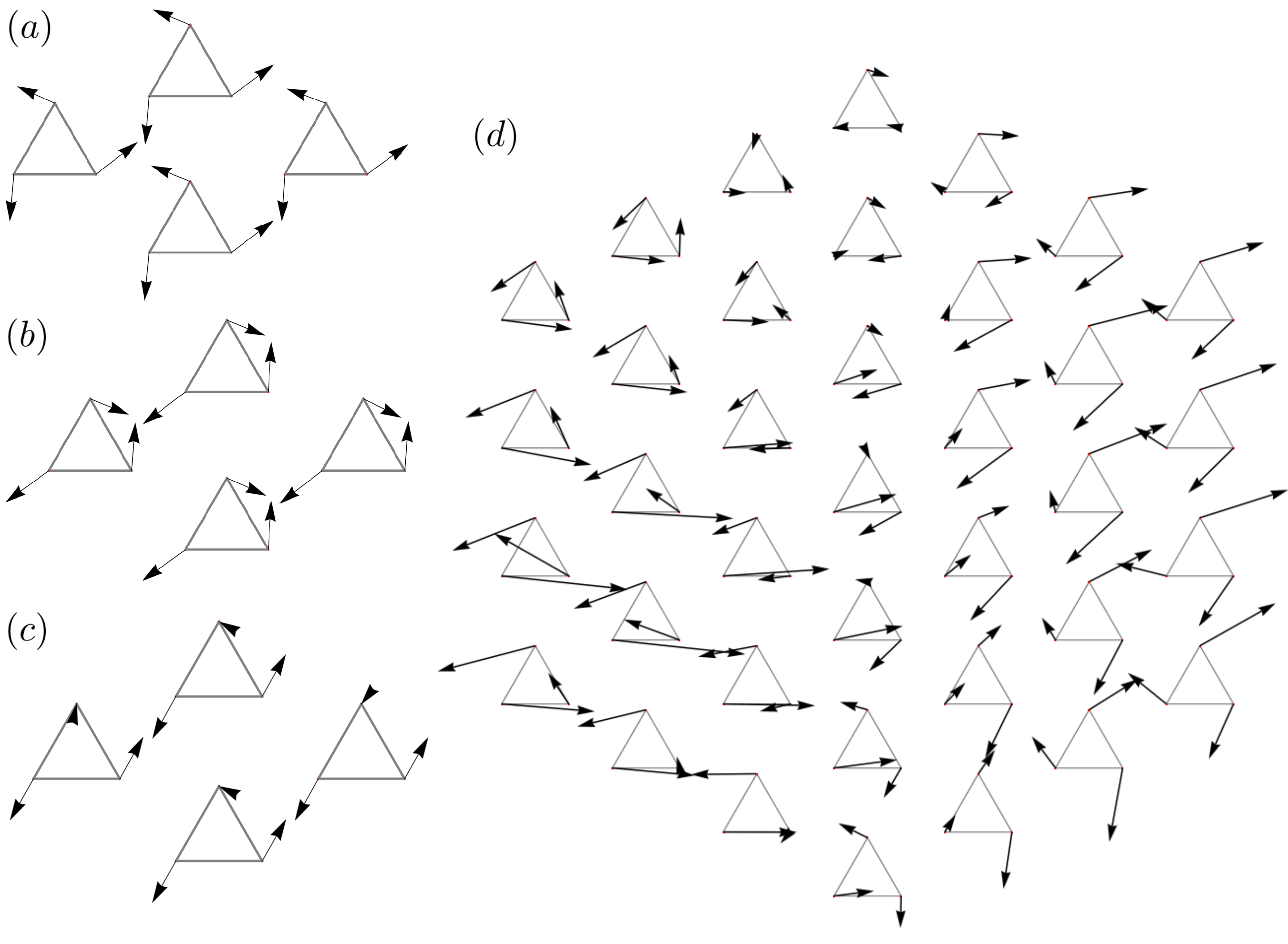}
	\vspace{-0.7cm}
	\caption{Individual stochastic realizations of atomic positions resulting in (a,b) spontaneous symmetry breaking in the atomic out-of-plane polarization density due to light scattering, (c) restored symmetry of unbroken state in ensemble-averaging, and (d) domain formation of two coexisting phases.  
    Length (angle) corresponds to dipole magnitude (phase). 
	(a,b) Small position uncertainty ($\sigma/a = 0.005$) randomly selects one of the modes in SSB. (c) The restored symmetry after the ensemble-average of $1600$ realizations. 
	(d) Large uncertainty ($\sigma/a = 0.03$) results in the coexistence of both phases in different spatial regions and the formation of domains. 
	The parameters as in Fig.~\ref{Fig:SymmetryBreakingOP}, with the laser frequency detuning from the atomic resonance $\Delta/\gamma = -12.1$ and intensity $\sum_j I^{(j)}/NI_{\rm sat}=0.37$.
    }
	\label{Fig:Disorder}
\end{figure}

In Fig.~\ref{Fig:Disorder}(a,b), we show how weak position fluctuations ($\sigma/a \alt 0.01$) act as a noise that randomly determines the dominantly occupied mode when the symmetry breaks. 
While any particular configuration of atom positions breaks the symmetry, we find that ensemble-averaging the response over many stochastic realizations restores the symmetry of the system, returning it to the unbroken state~\cite{Lee14}, as shown in Fig.~\ref{Fig:Disorder}(c). 
The out-of-plane polarization for the central unit cell of the array takes the value 
$\langle\textbf{b}^{x}\rangle=(5+8\text{i},-0.024+0.012\text{i},-5-8\text{i})^T\times10^{-2}$, with almost a complete loss of the polarization of the top atom, and equal magnitude polarization in opposite directions for the remaining two atoms, indicating the restored symmetry.
The average response is therefore markedly different to the response for any realization, e.g., Fig.~\ref{Fig:Disorder}(a,b), where the polarization of the top atom of each unit cell is comparable to one of its neighbors.
For large fluctuations ($\sigma/a\agt 0.01$), the symmetry breaks differently across the lattice, leading to the formation of domains in the collective excitation. This is shown in Fig.~\ref{Fig:Disorder}(d) for a particular realization of atom positions where $\textbf{b}^{x} \approx \bar{\textbf{v}}_{2}$ for each unit cell at $y\alt 0$, while $\textbf{b}^{x} \approx \bar{\textbf{v}}_{1}$ for unit cells with $y\agt 0$. Around $y\simeq 0$, the two domains smoothly connect, with an almost equal superposition of $\textbf{v}_{1}$ and $\textbf{v}_{2}$, and the symmetry unbroken.
Other realizations of the atom positions result in regions with almost no out-of-plane excitation mixed with regions where the symmetry is broken or unbroken. 
In all cases, domains smoothly connect, while domain size can vary from around ten unit cells to over half the lattice. 
There is also a decrease in the magnitude of the out-of-plane polarization, e.g., the average dipole $x$-component in Fig.~\ref{Fig:Disorder}(d) is approximately three times smaller than when the atoms are fixed at the lattice points, further decreasing for stronger fluctuations.

It is interesting to compare our stochastic electrodynamics simulations to standard quantum trajectories~\cite{dalibard1992,Tian92,Dum92} of continuously monitored open quantum systems. While a quantum master equation may describe an ensemble-averaged outcome of quantum measurement processes, quantum trajectories can reveal a possible measurement record of an individual experimental run. The detection record determines each quantum trajectory evolution where dynamics are conditioned on the previous measurement outcomes. Although the stochastic electrodynamics simulations do not concern the dynamical measurement process, the initially uncertain atomic positions are localized by the scattered light in each realization, representing one possible position configuration according to the quantum distribution. The atom array and light form an open dissipative system, where the position fluctuations result in both coherent and incoherent scattering~\cite{Bettles2020}. Consequently, there exists a nontrivial relationship between the geometry of the system and how accurately a scattered light measurement can localize the atomic positions. However, due to the strong nonlinearity, we find that even weak information about the atomic positions is sufficient to break the symmetry, resulting in measurement-induced symmetry breaking and phase transitions where each configuration is stochastically selected from the distribution given by quantum wave function of the atoms (compare with the cavity case of Ref.~\cite{Lee14}).

The SSB and dynamics we find here are very similar to those that occur in magnetic systems, and also in simulations of magnetism realized using superfluids in a standard triangular lattice~\cite{Struck2019}. 
The out-of-plane modes can be considered optical analogs of the phases in a classical XY model, with $\textbf{v}_{\Gamma;{\rm un}}$ identified as a ferromagnetic state, while the linear combinations ${\bf v}_{\rm ch1,ch2} \propto -(1\mp\text{i}){\bf v}_{\Gamma;1} + (1\pm\text{i}){\bf v}_{\Gamma;2}$ correspond to the degenerate configurations of the spiral state with uniform repetitions of the single unit-cell
\begin{align}\label{Eq:ChiralStates}
&\bar{{\bf v}}_{\rm ch1} \propto \begin{pmatrix} -1-\text{i} \\ 2 \\-1+\text{i},\end{pmatrix}, \,
\bar{{\bf v}}_{\rm ch2} \propto \begin{pmatrix} -1+\text{i} \\ 2 \\-1-\text{i},\end{pmatrix}.
\end{align}
When the real and imaginary parts of the polarization $x$-component are mapped to spin vectors ${\bf S}_{\rm ch1,ch2}^{(j)} = (\text{Re}[v^{x,(j)}_{\rm ch1,ch2}],\text{Im}[v^{x,(j)}_{\rm ch1,ch2}],0)^T$, the degenerate configurations can be distinguished by a chirality order parameter~\cite{Struck2019}
\begin{equation}\label{Eq:Chirality}
\kappa = \text{sgn}[{\bf S}_1\times{\bf S}_2+{\bf S}_2\times{\bf S}_3+{\bf S}_3\times{\bf S}_1],
\end{equation}
defined for the three atoms of a unit cell, where ${\bf v}_{\rm ch1}$ (${\bf v}_{\rm ch2}$) has $\kappa = +1$ ($\kappa = -1$).
The fact that similar dynamics and excitations can be observed in magnetic systems, superfluids and dipole-coupled atoms in triangular geometries is surprising given the substantial difference in the underlying physics of these systems. 
In Ref.~\cite{Struck2019}, the effective spin excitation is formed from a superfluid of Bose-condensed atoms with contact interactions, where the local superfluid phase at each lattice site maps to a classical spin.
SSB is then achieved by tuning the nearest-neighbor hopping between sites via shaking the lattice to induce a phase transition from the ferromagnetic to the degenerate spiral state, where one of the chiral configurations is randomly chosen.
In our system, however, trapped atoms are coupled via light-induced long-range dipole-dipole interactions, where the polarization density $x$-component acts as a classical spin, and all possible magnetic phases are accessible by appropriately exciting the corresponding modes. 
SSB is triggered by equally exciting the ${\bf v}_{1}$ and ${\bf v}_{2}$ modes instead of a ferromagnetic state, and then increasing the light intensity. In the symmetry-broken state, one of the ${\bf v}_{1}$, ${\bf v}_{2}$ mode populations, corresponding to one of the chiral states, then randomly dominates over the other. 
Collective optical excitations of atoms therefore offer a fascinating model for frustrated magnetic systems and SSB, and possibly other magnetic phenomena.

\begin{figure}
	\hspace*{-0cm}
	\includegraphics[width=0.9\columnwidth]{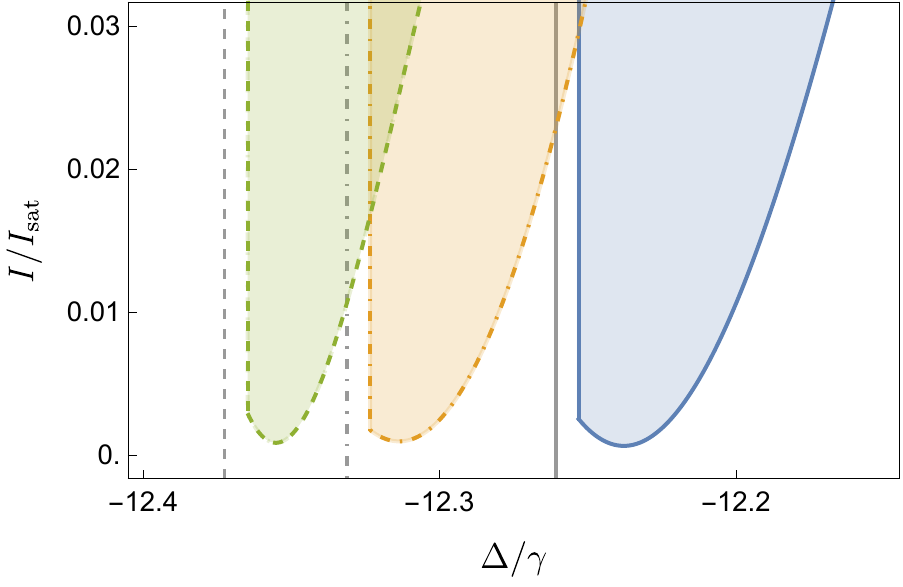}
	\vspace{-0.5cm}
	\caption{Diagram of spontaneously broken phases with light intensity and frequency thresholds in a driven atomic array, obtained from Eqs.~\eqref{Eq:ThreeModeModel}.
	Within certain ranges of the laser frequency detuning from the atomic resonance $\Delta$ and the intensity, $I/I_{\rm sat}$, SSB occurs in an array with $N=273$ (blue solid line), $N=507$ (orange dot-dashed line) or $N=993$ (green dashed line) atoms when $\Delta$ is greater than the ${\bf v}_{\Gamma;1}$ and ${\bf v}_{\Gamma;2}$ eigenmode resonance (corresponding black line for each $N$). 
    The parameters as in Fig.~\ref{Fig:SymmetryBreakingOP}.
	}
	\label{Fig:Nscaling}
\end{figure}

Our system also shares many similarities with optically-induced self-organization~\cite{Domokos2002,Black03,Asboth2005,Baumann2010,Niedenzu2013,Baumann2010,Labeyrie2014,Lee14,Robb15,Caballero-Benitez2015,Ivanov2020NAT,Baio21}, where small fluctuations seed symmetry-broken states. 
In particular, atom clouds in cavities or with optomechanical back-action only undergo self-organization and SSB for certain light frequencies and intensities. 
In Fig.~\ref{Fig:Nscaling}, we show analogous behavior by calculating the phase diagram based on the few-mode effective model, which demonstrates how SSB only occurs when the intensity is greater than a detuning-dependent threshold.
This intensity threshold takes minimal values when $\Delta \approx -\delta_d$, but for $\Delta < -\delta_d$, the symmetry cannot break for any incident field intensity, as found in the finite-size system numerics, Fig.~\ref{Fig:SymmetryBreakingOP}(c).
For increasing atom number, where the behavior for different $N$ is incorporated by using the finite-size system values for $\upsilon_d$ and $\delta_d$, the SSB region narrows and shifts as $\delta_d$ changes, while for different lattice spacings, the symmetry remains unbroken for any values of the detuning, intensity or level shift strength when $a\gtrsim 0.18\lambda$.
The requirement of small spacing is analogous to the emergence other strong nonlinear in dipole-coupled atoms, e.g., optical bistability and phase transitions~\cite{Parmee2018,Parmee2020,Parmee2021}, and dipole blockade~\cite{williamson2020b,Cidrim20}. 

The effective model allows for simple analytic expressions for the detuning threshold of SSB in Fig.~\ref{Fig:Nscaling} according to the coefficients of Eq.~\eqref{Eq:PQuadratic} (Appendix D), with the symmetry unbroken at any incident field intensity when
\begin{equation}\label{Eq:DetuningThresholdFull}
	\Delta > \Delta_{\text{th}} = (1-2P_{-})\left(-\delta_d + \frac{\upsilon_{d}(\gamma-\upsilon_d)}{\delta_d}\right).
\end{equation}
When $\Delta \approx -\delta_d$, low intensities are required for SSB, resulting in $\rho_{xx}^{(j)} \ll 1$ and the detuning threshold simplifying to $\Delta_{\rm th}\approx -\delta_d$, which agrees with the finite-size lattice results, Fig.~\ref{Fig:SymmetryBreakingOP}(c). 
The SSB intensity threshold can also be obtained in this limit by setting $\Delta = \Delta_{\rm th}+\epsilon$, where $\epsilon > 0$ and $\epsilon,\upsilon_d \ll \gamma, \alpha$, with
	\begin{align}\label{Eq:IntensityThresholdLimit}
		\frac{I}{I_{\text{sat}}} \gtrsim \frac{\upsilon_d^2\alpha^2\left[\gamma^2+\delta_d^2\right]}{\gamma^2\delta_d^3\epsilon}.
	\end{align}
The threshold scales with $\upsilon_d^2$, demonstrating that small intensities are needed to generate nonlinear effects in highly-subradiant modes~\cite{Williamson2020}, and also with $\alpha^2$ as the $J'=\pm1$ levels become shifted off-resonance for strong level shifts.
However, for $\alpha \rightarrow 0$, Eq.~\eqref{Eq:IntensityThresholdLimit} is no longer valid and the threshold intensity instead scales with $1/\alpha^2$ [see Eq.~\eqref{Eq:IntensityThresholdSimple} in Appendix D] as stronger intensities are required to generate nonlinearities when the coupling between the in-plane and out-of-plane modes decreases.

\begin{figure}
	\hspace*{-0cm}
	\includegraphics[width=\widthscale\columnwidth]{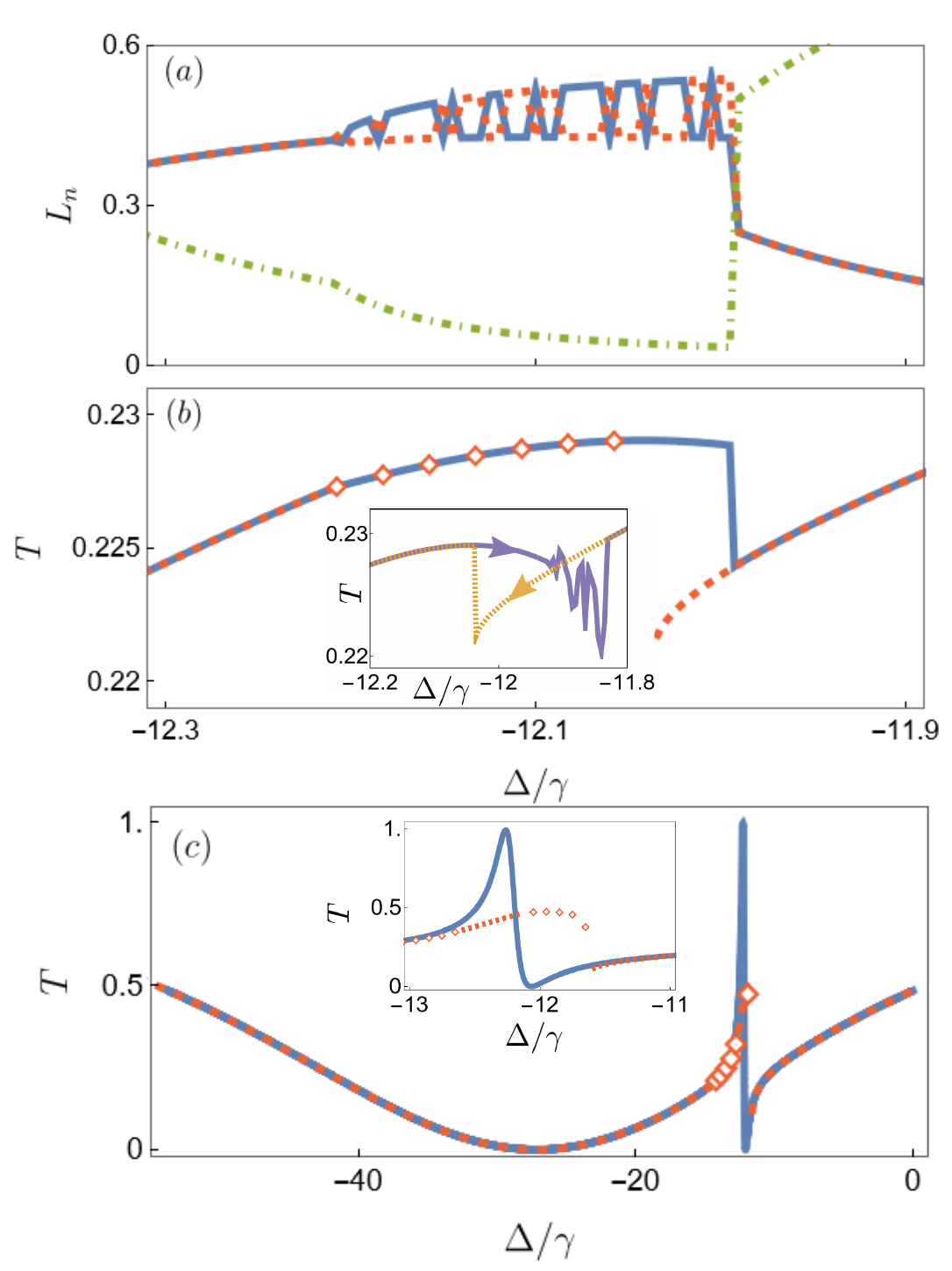}
	\vspace{-0.7cm}
	\caption{ 
	Coherent transmission and spontaneous symmetry breaking in the few-mode effective model [Eqs.~\eqref{Eq:ThreeModeModel}] for a large standard triangular array of atoms driven by light. 
	(a) Occupations $L_n$ [Eq.~\eqref{Eq:ModePopulation}] of ${\bf v}_{\Gamma;1}$ (blue line), ${\bf v}_{\Gamma;2}$ (red dashed line) and ${\bf v}_{\Gamma;y}$ (green dot-dashed line) demonstrating SSB at an intensity $\sum_j I^{(j)}/NI_{\rm{sat}}=3.3$, with the dominant mode switching due to noise as the laser frequency detuning from the atomic resonance $\Delta$ is varied.  
	(b) Coherent transmission $T$ [Eq.~\eqref{Eq:TransmissionAmplitude}] of the dynamics in (a) (blue line) and the dynamics under initial conditions which lead to the symmetry-unbroken steady state (red dashed line). When this state is unstable, the steady state instead has broken symmetry (red diamonds). 
	Transmission hysteresis (inset) occurs under a laser frequency sweep from $\Delta/\gamma = -13$ (purple right arrow) and $\Delta/\gamma = -10$ (orange left arrow).
    (c) Change in $T$ with intensity (inset shows a close up near the ${\bf v}_{\Gamma;1}$ and ${\bf v}_{\Gamma;2}$ resonance), where a sharp Fano resonance in the LLI (blue line) becomes a transmission jump (red dashed line and diamonds) at $\sum_j I^{(j)}/NI_{\rm{sat}}=0.6$ due to a similar steady-state change to (b). 
	The parameters as in Fig.~\ref{Fig:SymmetryBreakingOP}, except $\alpha/\gamma = -1.71$ in (c).
	}
	\label{Fig:Transmission}
\end{figure}

We find in Fig.~\ref{Fig:Transmission} how SSB results in a clear change in the coherent transmission $T$ (Appendix E) as the uniform in-plane mode population is changed in Eq.~\eqref{Eq:CSolnsMain}. 
However, $c_{y}$ depends only on $P_m^2$, and hence the coherent transmission is insensitive to which way the symmetry is broken (unlike incoherently scattered light which is sensitive to how the symmetry is broken). 
This is shown in Fig.~\ref{Fig:Transmission}(b) as a smooth transmission curve, while the mode occupations in Fig.~\ref{Fig:Transmission}(a) exhibit random jumps due to noise changing the dominant mode. 
The presence of different steady states in the dynamics also leads to hysteresis of the transmission when sweeping the laser frequency [the inset of Fig.~\ref{Fig:Transmission}(b)]. 
Due to the insensitivity of $T$ on how the symmetry breaks, even though there are three steady-state solutions, there is a coexistence of only two transmission curves which depend on the starting steady state before the sweep, similar to the behavior seen for optical bistability.
The spikes in the transmission indicate the emergence of an additional symmetry-broken solution.
The transmission from the array also demonstrates EIT-like behavior with a Fano resonance [Fig.~\ref{Fig:Transmission}(c)] that becomes particularly clear in the limit of LLI, where $T\simeq 1$ when $\Delta = -\delta_d$ provided that $\alpha^2 \gg \upsilon_{y}\upsilon_{d}$. 
Unlike for the independent-atom case, the Fano resonance width is not limited by the single-atom linewidth $\gamma$, but by the much narrower $\upsilon_d$. 
At stronger $\alpha$, the width broadens to $(w_{+}+w_{-})/2$ where $w_{\pm}=(\delta_d-\delta_y\pm \upsilon_{y})\mp\sqrt{(\delta_d-\delta_y\pm \upsilon_{y})^2+4\alpha^2}$.
Increasing the incident field intensity results in the resonance transforming into a jump in transmission where the dynamics changes from a symmetry-broken steady state to an unbroken one. There is also a loss of complete transmission because of linewidth power broadening, which results in a reduction in the transfer of the in-plane to out-of-plane population. However, for larger $\alpha$, full transmission would be recovered.

\subsection{In-plane spontaneous symmetry breaking}
\label{sec:ip_symmetry}

\begin{figure}
	\hspace*{-0cm}
	\includegraphics[width=\columnwidth]{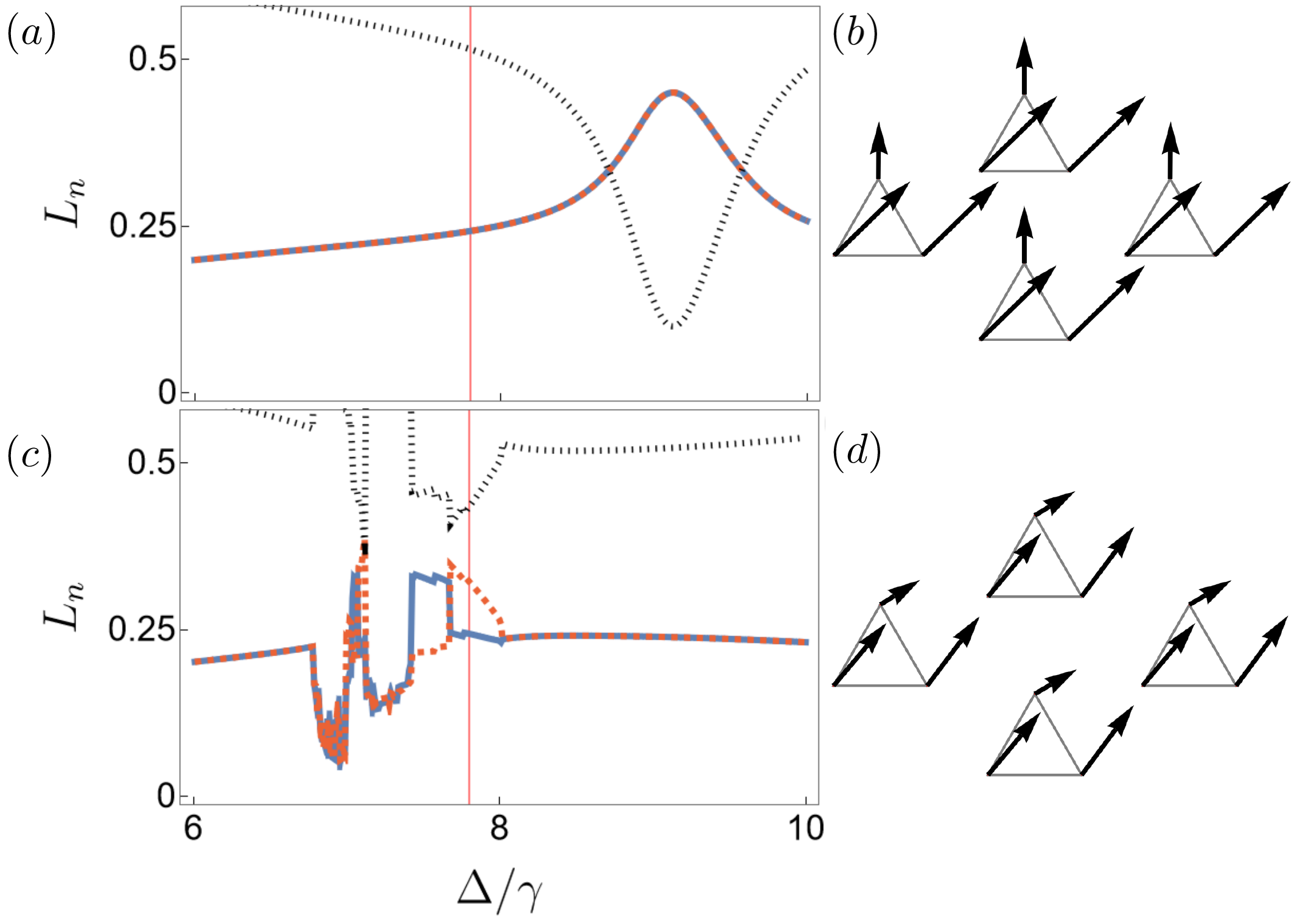}
	\vspace{-0.7cm}
	\caption{
	Spontaneous symmetry breaking shown by the atomic in-plane polarization density and mode occupation spectrum in an array of $N=273$ atoms driven by a $z$-polarized plane wave, with intra- and inter-unit-cell spacings $a=0.1\lambda$ and $s=6\sqrt{3}a$, respectively.
	(a,c) Occupations $L_n$ [Eq.~\eqref{Eq:ModePopulation}] of ${\bf v}_{a}$ (blue line) and ${\bf v}_{b}$ (red dashed line, and the sum of other eigenmode occupations (black dotted line) as a function of the laser frequency detuning from the atomic resonance $\Delta$. (a) Unbroken symmetry with equal populations in the LLI limit, while for the intensity (c) $\sum_j I^{(j)}/NI_{\rm{sat}}=0.6$, the mode symmetry is broken with a corresponding change in the polarization $y$- and $z$-components (component magnitudes projected onto $y$ and $z$ axes) (b,d) shown for $\Delta/\gamma=7.8$ (red line).
	}
	\label{Fig:SymmetryBreakingIP}
\end{figure}

SSB can also be observed for excitations for which the dipole moments are oriented in the lattice plane. These modes can be directly driven by the incident field without externally imposed level shifts. 
We drive the array with a $z$-polarized plane-wave, equally populating the degenerate eigenmode pair $\textbf{v}_{a}$ and $\textbf{v}_{b}$, which are the corresponding finite-lattice versions of $\textbf{v}_{\Gamma;a}$ and $\textbf{v}_{\Gamma;b}$ discussed in Sec.~\ref{KyleLLI}, and consider an array with $s$ notably larger than $\sqrt{3}a$ as the coupling between the incident field and $\textbf{v}_{a}$, $\textbf{v}_{b}$ vanishes for the standard triangular lattice.
In the LLI limit, the mode occupations are symmetric, $L_a=L_b$ [Fig.~\ref{Fig:SymmetryBreakingIP}(a)], and in the vicinity of the eigenmode resonance, $\Delta/\gamma \simeq 9.1$, the population of other modes is small, with maximal values $L_{a,b} \simeq 0.45$ at the resonance itself. 
Similarly to the out-of-plane mode case, the mode symmetry can be determined by the atomic polarization density. An equal superposition of ${\bf v}_{a}$ and ${\bf v}_{b}$ results in a purely radial dipole moment on the top atom in each unit cell in the steady-state excitation [Fig.~\ref{Fig:SymmetryBreakingIP}(b)]. 

Upon increasing the incident field intensity, SSB occurs [Fig.~\ref{Fig:SymmetryBreakingIP}(c)], signaled by the polarization of the top atom in each unit cell obtaining a nonzero tangential component [Fig.~\ref{Fig:SymmetryBreakingIP}(d)], and a corresponding change in the mode occupations with $L_a \neq L_b$.
The symmetry breaking is spontaneous as also indicated by small changes to the initial conditions in the dynamics switching the dominant mode [from ${\bf b}=0$ to ${\bf b} \simeq 0.001{\bf v}_{a}$ (${\bf b} \simeq 0.001{\bf v}_{b}$) resulting in $L_{a}>L_{b}$ ($L_{b}>L_{a}$)]. 
While changing the initial conditions seeds the final steady-state, fluctuations in the atom positions do not, and we find that even small position changes ($\sigma/a \agt 0.005$) break the uniformity of the collective excitation. Rather than leading to the formation of different symmetry broken domains, as was the case for the out-of-plane modes, the dipole orientations on different unit cells instead vary rapidly and seemingly randomly across the lattice. The in-plane SSB is therefore not similarly robust to position fluctuations as the out-of-plane SSB.

\section{Concluding remarks}

We analyzed the semiclassical behavior of densely packed atoms on a triangular array of three-atom unit cells, where frustration emerges due to competition of the dipolar interactions and the lattice geometry.
In the limit of LLI, we found the frustration leads to the formation of degenerate pairs of the collective LLI eigenmodes for both a single isolated triangle and arrays of triangles. The system exhibits many close to degenerate eigenmodes that are reflected in a nearly flat band structure. 

We showed how to engineer steady-state collective radiative excitations with an equal population of two degenerate modes. 
Upon increasing the intensity of the incident field, we found this mode occupation symmetry is spontaneously broken, with the system randomly choosing one mode over the other. Fluctuations in the atomic position seed the SSB and could be interpreted as a light-induced measurement of atomic positions from a distribution that is determined by the quantum wave function. The measurement-induced symmetry breaking and corresponding phase transition lead to uniform symmetry broken phases for small position uncertainties. In contrast, larger uncertainties lead to the formation of domains where the symmetry is broken differently across the lattice. Ensemble-averaging over many stochastic realizations restores the symmetry. 

The SSB of the degenerate out-of-plane modes can be considered an optical analog of frustrated magnetism. The two degenerate modes correspond to the spiral ground state configurations in a closed magnetic system. The SSB we find is also analogous to pattern formation and self-organization in cold gas clouds, where fluctuations in the atomic positions seed the symmetry broken state and where SSB can only occur above an intensity threshold.
Therefore, dipole-coupled atoms could be utilized to simulate magnetic systems and achieve optically controlled symmetry breaking. In the absence of decoherence, the cooperative interactions can reach the strongly-coupled quantum regime~\cite{williamson2020b}, providing a possibility to simulate quantum magnetism without the need to reach ultralow temperatures~\cite{Bloch08}.

Data used in the publication is available at (DOI TO BE ADDED IN PROOF).

\section{Acknowledgments}
We acknowledge financial support from the UK EPSRC (Grant Nos.\ EP/S002952/1, EP/P026133/1, and EP/W005638/1).

\begin{appendices}

\section{Stochastic electrodynamics}

To allow for position fluctuations of the atoms, a set of fixed positions $\{ {\bf r}_1, \ldots, {\bf r}_N\}$ are sampled from a joint probability distribution $P({\bf r}_1, \ldots, {\bf r}_N)$ for each stochastic realization. The probability distribution is given by the absolute square of the many-body wavefunction, and ensemble-averaging over many stochastic realizations yields the expectation values of the observables~\cite{Javanainen1999a,Lee16,Bettles2020}.
We assume that the atoms are trapped with one atom per lattice site in the vibrational ground states of the individual sites, with root-mean-square width $\sigma$. 
The sampling of such a Mott-insulator ground state then simplifies to the sampling of an independent atom in each site~\cite{Jenkins2012a}.

Even though we factorize quantum correlations between atoms to obtain Eqs.~\eqref{Eq:SpinEquations}, significant light-induced classical correlations between atoms at positions ${\bf r}$ and ${\bf r}'$ are still present due to the dipole-dipole interactions. This means in terms of correlation functions that, in general, $$\langle \hat{\psi}^{\dagger}_a({\bf r})\hat{\psi}^{\dagger}_b({\bf r}')\hat{\psi}_c({\bf r}')\hat{\psi}_d({\bf r}) \rangle \neq \langle\hat{\psi}^{\dagger}_a({\bf r})\hat{\psi}_d({\bf r})\rangle\langle\hat{\psi}^{\dagger}_b({\bf r}')\hat{\psi}_c({\bf r}')\rangle,$$ where $a,b,c,d\in g,\mu$ and the atomic field operators for the ground and excited states are denoted by $\hat{\psi}_{g,\mu}({\bf r})$, such that $\rho_{g\mu}^{(j)} = \langle\hat{\psi}_{g}^{\dagger}({\bf r}_j)\hat{\psi}_{\mu}({\bf r}_j)\rangle$ and $\rho_{\mu\nu}^{(j)} = \langle\hat{\psi}_{\mu}^{\dagger}({\bf r}_j)\hat{\psi}_{\nu}({\bf r}_j)\rangle$, etc.
The light-induced correlations can markedly change the optical response in the presence of position fluctuations~\cite{Javanainen2014a,JavanainenMFT}.

For any position configuration $\{ {\bf r}_1, \ldots, {\bf r}_N\}$, we calculate the total field at each atom 
$\textbf{E}{}^+(\textbf{r}) =  \boldsymbol{\mathcal{E}}{}^+(\textbf{r})+\textbf{E}{}^+_s(\textbf{r})$, where the scattered light
\begin{equation} \label{Efieldscattered}
\epsilon_0\textbf{E}{}^+_{s}(\textbf{r}) =\sum_j\mathsf{G}(\textbf{r}-\textbf{r}_j)\textbf{d}_j.
\end{equation}
$\textbf{E}{}^+(\textbf{r})$ satisfies Maxwell's wave equation with an atomic polarization source~\cite{Ruostekoski1997a}, where the dipole radiation kernel acting on a dipole located at the origin, with $r= |\textbf{r}|$ and $\hat{\mathbf{r}}=\textbf{r}/r$, is given by the familiar formula~\cite{Jackson,BOR99}
\begin{align}\label{Gdef}
\mathsf{G}(\mathbf{r})\mathbf{d}&=-\frac{\mathbf{d}\delta(\mathbf{r})}{3}+\frac{k^3}{4\pi}\Bigg\{\left(\hat{\mathbf{r}}\times\mathbf{d}\right)\times\hat{\mathbf{r}}\frac{e^{ikr}}{kr}\nonumber\\
&\phantom{=}-\left[3\hat{\mathbf{r}}\left(\hat{\mathbf{r}}\cdot\mathbf{d}\right)-\mathbf{d}\right]\left[\frac{i}{(kr)^2}-\frac{1}{(kr)^3}\right]e^{ikr}\Bigg\}.
\end{align}

\section{Triangle eigenmodes}
For in-plane eigenmodes of an isolated single triangle, it is convenient to work in a radial and tangential basis, $\hat{\textbf{e}}_{r} = [\cos(\phi)\hat{\textbf{e}}_y + \sin(\phi)\hat{\textbf{e}}_z]/\sqrt{2}$ and $\hat{\textbf{e}}_{t} = [-\sin(\phi)\hat{\textbf{e}}_y + \cos(\phi)\hat{\textbf{e}}_z]/\sqrt{2}$ where $\phi$ is the polar angle defined with respect to the $y$-axis. 
The degenerate in-plane eigenmodes in Eqs.~\eqref{Eq:EigenmodesinPlaneAB} then depend on the triangle size via $\zeta_{\pm}$ where, 
\begin{align}\label{Eq:alpha}
		&\zeta_{\pm} = \pm\frac{-6+6\text{i}ka+2(ka)^2}{3\sqrt{3}[1-\text{i}ka+(ka)^2]}\pm\nonumber\\
		&\frac{\sqrt{45-90\text{i}ka-51(ka)^2+6\text{i}(ka)^3+13(ka)^4}}{3\sqrt{3}[1-\text{i}ka+(ka)^2]}.
\end{align}

\section{Coupling to out-of-plane modes}

The polarization of $\textbf{v}_{1}$ and $\textbf{v}_{2}$ [Eq.~\eqref{Eq:Eigenmodes}] lies perpendicular to the lattice plane, and so they do not couple to the incident light directly. 
Instead, the incident light couples to the in-plane mode $\textbf{v}_y$ which, when the $m_{J'}=\pm1$ levels are degenerate, evolves independently with no coupling to other modes. 
When the level shifts do not average to zero [$\delta_{+}^{(j)}+\delta_{-}^{(j)} \neq 0$], however, $\textbf{v}_y$, $\textbf{v}_1$, and $\textbf{v}_2$ are no longer eigenmodes of the full LLI evolution matrix $\mathcal{H}+\delta\mathcal{H}$ in Eq.~\eqref{LLI}, and are instead coupled together (as, in general, are other modes). Then it is possible to excite the out-of-plane modes 
by first driving $\textbf{v}_y$ with the incident field and using the varying level shifts to resonantly couple to $\textbf{v}_1$ and $\textbf{v}_2$, transferring the excitation~\cite{Facchinetti16,ballantine2020}.

To excite both out-of-plane modes equally, we target a symmetric combination $\textbf{v}_{+}$, where we introduce the orthonormal symmetric and antisymmetric combinations ${\bf v}_{\pm} = \mathcal{N}_{\pm}({\bf v}_{1}\pm {\bf v}_{2})$ with normalization constants $\mathcal{N}_{-}=1/\sqrt{3}$, $\mathcal{N}_{+}=1$.  
In particular, $\textbf{v}_+$ corresponds to an approximately uniform repetition of the single unit-cell combination
\begin{equation}
\bar{\textbf{v}}_{+}^x = \bar{\textbf{v}}_{1}^x+\bar{\textbf{v}}_{2}^x =\frac{1}{\sqrt{2}}\begin{pmatrix} -1 \\ 0 \\ +1 \end{pmatrix}, \quad \bar{\textbf{v}}_{+}^{r},\bar{\textbf{v}}_{+}^{t}=0.
\end{equation}
The shifts on each atom couple the coherences $\rho_{gx}^{(j)}$ and $\rho_{gy}^{(j)}$, with $\delta_{\pm}^{(j)}$ having the same sign resulting in an out-of-plane polarization component with the same magnitude but a $\pi$ phase difference. 
We therefore choose shifts
\begin{align}\label{Eq:ZeemanShift}
&\delta_{\pm}^{(j)} = \alpha \sqrt{\frac{3}{2}} \begin{pmatrix} -1 \\ 0 \\ +1 \end{pmatrix},
\end{align}
on each unit cell to couple the uniform $\mathbf{v}_y$ to $\textbf{v}_{+}$, with tunable strength $\alpha$.

The desired spatial variation in the level shifts can be achieved by using the ac Stark shift~\cite{gerbier_pra_2006} of a standing wave laser, with the intensity varying across the atoms in a unit cell and periodic between different unit cells. 
For example, a control field $\cbE_c(\vec{r})=\mathcal{E}_{c}\unitvec{e}_{+} \cos{(\unitvec{k}_c\cdot\vec{r}+\phi)}$, with $k_{c,y}=\pi/(3 a)$, $k_{c,z}=\pi/(\sqrt{3}a)$, and $\phi=\pi/4$, and where $\vec{r}=0$ is the position of an atom at the top of a unit cell, has periodic intensity between different unit cells and an equal intensity gradient between atoms one and two, and atoms two and three. The circular polarization induces a vector polarizability which splits the $m_{J'}=\pm 1$ levels~\cite{Schmidt16,Rosenbusch09,LeKien13}. For appropriately chosen intensity and detuning from an off-resonant transition, this leads to level shifts of the form 
\begin{equation}
\delta_{\pm}^{(j)}= \alpha\sqrt{\frac{3}{2}}
\begin{pmatrix}0 \\  1 \\ 2\end{pmatrix},
\end{equation}
up to an arbitrary overall shift depending on the details of the transition.
Then an overall linear Zeeman splitting from a magnetic field or uniform-intensity laser or microwave chosen to provide an atom-independent shift $\delta_{\pm}^{(j)}=-\alpha \sqrt{3/2}$ gives the desired level shifts in Eq.~(\ref{Eq:ZeemanShift}). 
Alternatively, atoms at different lattice sites could be prepared in different hyperfine states~\cite{mandel03}, with the atomic transitions blue-detuned, on resonance or red-detuned across the three atoms of the unit cell.

The transfer of population from the directly driven in-plane modes to the subradiant out-of-plane modes is reminiscent of EIT in noninteracting atoms~\cite{FleischhauerEtAlRMP2005}. Here, $\textbf{v}_y$ plays the role of a `bright' state, which couples directly to the incident field, while the linear combination $\textbf{v}_{+}$ plays the role of a `dark' state, which does not. It is the highly subradiant nature of the out-of-plane modes ($\upsilon_1,\upsilon_2\ll \gamma$, see Sec.~\ref{KyleLLI}), which arises due to collective many-atom effects, that allows the efficient occupation of $\textbf{v}_{+}$, even for small $\alpha$.

\section{Effective few-mode model}

In the large lattice limit, the population of eigenmodes not targeted by the level shifts Eqs.~\eqref{Eq:ZeemanShift} becomes negligible, as too does any in-plane mode except ${\bf v}_{\Gamma;y}$ due to phase-matching. 
The atomic polarization density is then approximately ${\bf b} = \sqrt{N}[c_1{\bf v}_{\Gamma;1}+c_2{\bf v}_{\Gamma;2}+c_{y}{\bf v}_{\Gamma;y}]$, and near the $\textbf{v}_{\Gamma;1}$ and $\textbf{v}_{\Gamma;2}$ resonance, the occupation of $\textbf{v}_{\Gamma;y}$ will be small, such that only $\rho_{xx}^{(j)}$ [Eq.~\eqref{Eq:Cartesian}] will be nonzero.
Using these approximations, the dynamics [Eqs.~\eqref{Eq:SpinEquations}] simplify to
\begin{subequations}\label{Eq:ThreeModeModel}
	\begin{align}
	\dot{c}_{\pm} &= \text{i}(\lambda_{d}+\Delta)c_{\pm}-\alpha_{\pm} c_{y}\nonumber\\
	&\phantom{===}-2\text{i}[(\lambda_{d}-\text{i}\gamma)c_{\pm}]P_{\pm}-2\text{i}[(\lambda_{d}-\text{i}\gamma)c_{\mp}]P_{m} , \label{Eq:ThreeModeModel:cpm}\\ 
	\dot{c}_{y} &= \text{i}(\lambda_{y}+\Delta)c_{y}+\alpha_+ c_+ \nonumber \\
	&\phantom{===}+\text{i}(1-P_t)\mathcal{R}_{gy}-\text{i}[(\lambda_{y}-\text{i}\gamma)c_{y}]P_t, \label{Eq:ThreeModeModel:cy}\\ 
	\dot{P}_{\pm} &= -2\gamma P_{\pm} +\frac{3}{2}\text{i}[(\lambda_d-\text{i}\gamma)c_{\pm}]c_{\pm}^* +\text{c.c.}\nonumber\\
	&\phantom{===}+\frac{1}{2}\text{i}[(\lambda_d-\text{i}\gamma)c_{\mp}]c_{\mp}^*+\text{c.c.},\label{Eq:ThreeModeModel:Ppm}\\
	\dot{P}_{m} &= -2\gamma P_{m} +\text{i}[(\lambda_d-\text{i}\gamma)c_{+}]c_-^*\nonumber\\
	&\phantom{===}+\text{i}[(\lambda_d-\text{i}\gamma)c_{-}]c_+^*+\text{c.c.},\label{Eq:ThreeModeModel:PM}
	\end{align}
\end{subequations}
where $c_{\pm} = (c_{1}\pm c_{2})/2\mathcal{N}_{\pm}$, $\alpha_+=\alpha$, $\alpha_-=0$, and collective terms in square brackets are obtained using 
\begin{subequations}\label{Eq:collective}
\begin{align}
& \sum_{\nu}\sum_{l \neq j}\frac{6\pi\gamma}{k^3}\hat{\textbf{e}}_{\mu}^*\cdot \left[\mathsf{G}(\textbf{r}_j-\textbf{r}_l)\hat{\textbf{e}}_{ \nu}\right]{\bf v}^{\nu,(l)}_{1,2}=(\lambda_{d}-\text{i}\gamma){\bf v}^{\mu,(j)}_{1,2},\\
& \sum_{\nu}\sum_{l \neq j}\frac{6\pi\gamma}{k^3}\hat{\textbf{e}}_{\mu}^*\cdot \left[\mathsf{G}(\textbf{r}_j-\textbf{r}_l)\hat{\textbf{e}}_{ \nu}\right]{\bf v}^{\nu,(l)}_{y}=(\lambda_{y}-\text{i}\gamma){\bf v}^{\mu,(j)}_{y},
\end{align}
\end{subequations}
but with the finite-size eigenmodes then approximated by their corresponding infinite counterparts.
Equations~\eqref{Eq:ThreeModeModel} are reminiscent to the dynamics of optical Bloch equations of an atom. The incident field $\mathcal{R}_{gy}$ drives the transition from the ground state to an intermediate state with coherence $c_{y}$, and linewidth $\upsilon_y$, while $\alpha_\pm$ represent coupling fields from this intermediate state to the final states with populations $P_\pm$, coherences $c_\pm$ and linewidth $\upsilon_d$.
For the symmetry-unbroken state where $c_-=0$, there are, therefore, close similarities with a three-level system and EIT. 
This analogy is even more accurate in the LLI limit where $P_{\pm}=0$, and where $\alpha_+$ acts as a coupling field and $\mathcal{R}_{gy}$ the probe field.
From the steady-state mode coefficients of Eqs.~\eqref{Eq:CSolnsMain}, $c_+/c_y = -\alpha/\tilde{\upsilon}_d$ at $\Delta=-\tilde{\delta}_d$, where $\tilde{\delta}_d$ ($\tilde{\upsilon}_d$) corresponds to the real (imaginary) part of Eq.~\eqref{Eq:Zdefnd}.
It is therefore easy to see in the LLI limit how the population is transferred to the out-of-plane mode when $\alpha \gg \upsilon_d$, resulting in a Fano resonance at $\Delta=-\delta_d$ where the transmission is enhanced. However, beyond the LLI limit, this resonance is shifted and transmission gets reduced due to broadening from nonlinearity where $\tilde{\upsilon}_d\sim \alpha$, as shown in Fig.~\ref{Fig:Transmission}(c).

There are symmetry-broken solutions to Eqs.~\eqref{Eq:ThreeModeModel} only in a small window of frequencies. The frequencies where $P_m \neq 0$ can be determined by substituting the mode coefficients of Eqs.~\eqref{Eq:CSolnsMain} into Eqs.~\eqref{Eq:ThreeModeModel:PM}, giving  Eq.~\eqref{Eq:PQuadratic} with coefficients,
\begin{subequations}\label{Eq:QuadraticABC}
\begin{align}
A_1 =& 16\gamma |Z_{y}|^2|\lambda_{d}-\text{i}\gamma|^4,\\
A_2 =& 8\gamma \text{Re}[Z_-Z^*_{y}(\lambda_{d}^*+\text{i}\gamma)^2(\alpha^2-Z_{y}Z_+)],\\
A_3 =& \gamma|Z_-|^2|\alpha^2-Z_{y}Z_+|^2+\nonumber\\
&2\text{Im}[\lambda_d-\text{i}\gamma]\text{Re}[(\lambda_{d}-\text{i}\gamma)Z_-^*]\alpha^2\mathcal{R}_{gy}^2(P_t-1)^2.
\end{align}
\end{subequations}
Equations~\eqref{Eq:QuadraticABC} depend on $P_{\pm}$ [Eqs.~\eqref{Eq:Zdefn}], and so Eq.~\eqref{Eq:PQuadratic} has to be solved with Eqs.~\eqref{Eq:ThreeModeModel:Ppm} [using the mode coefficients of Eqs.~\eqref{Eq:CSolnsMain}] to obtain an explicit solution for $P_m$.  
However, analyzing Eq.~\eqref{Eq:PQuadratic} using Descartes' rule of signs gives the semianalytic detuning threshold Eq.~\eqref{Eq:DetuningThresholdFull} for a single nonzero value of $P_m^2$ to exist and for the symmetry to break. 
Similarly, the intensity threshold has a semianalytic form for $\Delta>\Delta_{\rm th}$,
\begin{equation}\label{Eq:IntensityThresholdSimple}
\frac{I}{I_{\text{sat}}} \gtrsim \frac{|Z_-|^2|\alpha^2-Z_+Z_y|^2}{\alpha^2\gamma(\gamma-\upsilon_d)(P_t-1)^2 \text{Re}[(\lambda_d-\text{i}\gamma)Z_-^*]},
\end{equation}
which simplifies to Eq.~\eqref{Eq:IntensityThresholdLimit} near $\Delta_{\rm th}$, as $P_{\pm}\approx0$.
Equation~\eqref{Eq:IntensityThresholdSimple} scales with $\alpha^2$ when $\alpha \gg \gamma$ and diverges in the limit $\alpha\rightarrow0$ as discussed in Sec.~\ref{sec:oop_symmetry}.

\section{Coherent transmission}

We calculate the coherent (power) transmission, $T=|t|^2$, for a large subwavelength 2D lattice where the amplitude
\begin{equation}\label{Eq:TransmissionAmplitude}
	t=\frac{\int\hat{\textbf{e}}\cdot\textbf{E}^+(\textbf{r})dS}{\int\hat{\textbf{e}}\cdot\boldsymbol{\mathcal{E}}{}^+(\textbf{r})dS}.
\end{equation}
Only the zeroth order Bragg peak exists, such that the light is scattered purely in the forward and backward direction. Therefore, for an excitation with a spatially uniform phase profile, the coherently transmitted light at a point $(x,0,0)$ can be approximated by~\cite{dalibardexp,Javanainen17,Facchinetti18,Javanainen19}
\begin{equation}\label{EFieldSlab}
		\textbf{E}{}^+(x)=\mathcal{E}_0\hat{\textbf{e}}e^{\text{i}kx} + \frac{\text{i}k}{2\mathcal{A}\epsilon_0}\sum_{l}^{}[\hat{\textbf{d}}_l-(\hat{\textbf{e}}_x\cdot\hat{\textbf{d}}_l)\hat{\textbf{e}}_x]e^{\text{i}kx},
\end{equation}
when $\lambda \lesssim x \ll \sqrt{\mathcal{A}} $, where $\mathcal{A}$ is the total area of the array.

\end{appendices}

%

\end{document}